\documentstyle[pre,aps,preprint,epsf,eqsecnum]{revtex} 

\tightenlines

\newcommand{\amax}{a^{\mathrm{max}}}

\newcommand{\app}{\rightarrow}

\newcommand{\asubpp}{{a_{++}}}
\newcommand{\asubpm}{{a_{+-}}}
\newcommand{\asubmm}{{a_{--}}}

\newcommand{\biguint}{U^{\mathrm{int}}}

\newcommand{\bigutot}{U^{\mathrm{tot}}}

\newcommand{\dee}{\partial}

\newcommand{\delm}{\delta_-}
\newcommand{\delp}{\delta_+}

\newcommand{\delsig}{\delta_\sigma}

\newcommand{\dhtext}{Debye-H\"{u}ckel\ }

\newcommand{\dtrstd}{\!\!{\mathrm d}^3 r \,}

\newcommand{\elcharge}{q_0}
\newcommand{\egenl}{{\cal F}}
\newcommand{\eless}{{\cal F}^<}
\newcommand{\egrtr}{{\cal F}^>}

\newcommand{\expmf}{EXP$^<$MF$^>$ADH\ }

\newcommand{\fbar}{\bar{f}}

\newcommand{\irm}{{\mathrm{I}}}
\newcommand{\iirm}{{\mathrm{II}}}
\newcommand{\iiirm}{{\mathrm{III}}}
\newcommand{\kapd}{\kappa_D}

\newcommand{\lbrak}{\left [}
\newcommand{\lpar}{\left (}

\newcommand{\potij}{\varphi_{ij}}
\newcommand{\pottwid}{\tilde{\varphi}}
\newcommand{\potijtwid}{\tilde{\varphi}_{ij}}
\newcommand{\psihat}{\hat{\psi}}

\newcommand{\rbf}{{\mathbf{r}}}
\newcommand{\rbrak}{\right ]}

\newcommand{\rhohat}{\hat{\rho}}
\newcommand{\rhostar}{\rho^*}

\newcommand{\rpar}{\right )}

\newcommand{\thsig}{\theta_\sigma}

\newcommand{\thp}{\theta_+}

\newcommand{\tstar}{T^*}

\newcommand{\uself}{u^{\mathrm{self}}}

\newcommand{\xd}{x_D}

\newcommand{\zm}{z_-}
\newcommand{\zp}{z_+}

\newcommand{\zsig}{z_\sigma}

\newcommand{\boxhalf}{\hbox{$\frac{1}{2}$}}
\newcommand{\threehalves}{\hbox{$\frac{3}{2}$}}

\newcommand{\boxthird}{\hbox{$\frac{1}{3}$}}
\newcommand{\twothirds}{\hbox{$\frac{2}{3}$}}

\begin{document}
%\draft
\title{Asymmetric Primitive-Model Electrolytes:\\ 
       Debye-H\"{u}ckel Theory, Criticality and Energy Bounds}
\author{Daniel M. Zuckerman\footnote{Current address: Department of Physiology, Johns Hopkins University, School of Medicine, 725 N. Wolfe St., Baltimore, MD 21205}, Michael E. Fisher, and Stefan Bekiranov\footnote{Current address: Laboratory of Computational Genomics, The Rockefeller University, New York, NY 10021}}
\address{Institute for Physical Science and Technology,\\ University of Maryland, College Park, Maryland 20742}
%\date{\today}
\maketitle

%\begin{center}
%\large Abstract
%\end{center}

\dhtext (DH) theory
is extended to treat two-component size- and charge-asymmetric primitive
models, focussing primarily on the 1:1 additive hard-sphere electrolyte
with, say, negative ion diameters, $\asubmm$, larger than the positive ion
diameters, $\asubpp$.  
The treatment highlights the crucial importance of
the charge-unbalanced ``border zones'' around each ion into which other
ions of only one species may penetrate.  
Extensions of the DH approach which describe
the border zones in a physically reasonable way are exact at high $T$ and low density, $\rho$, and, furthermore, are also in substantial
agreement with recent simulation predictions for \emph{trends} in the critical
parameters, $T_c$ and $\rho_c$, with increasing size asymmetry.  
Conversely,
the simplest linear asymmetric DH description, which fails to account for
physically expected behavior in the border zones at low $T$, can violate
a new lower bound on the energy (which applies generally to models asymmetric in
both charge and size).  
Other recent theories, including those based on
the mean spherical approximation, have predicted trends in the critical
parameters quite opposite to those established by the simulations.

\vspace*{0.3cm}
\noindent PACS numbers: 02.70.Lq, 05.70.Jk, 64.70.Fx

\newpage
\section{Introduction}
While the values of the critical parameters for the fully symmetric hard-sphere model of an electrolyte (the ``restricted primitive model'' or RPM \cite{debye,bjerrum,falkeb}) have received extensive theoretical attention for a number of years (see, e.g., \cite{falkeb,stellwularsen,gil,ebgrigo,fisher,stellrev,Fisher-Levin,critique}), only recently has interest focussed on the effects of \emph{asymmetry} on phase-coexistence and criticality \cite{sabir,Gonzalez-1999,Raineri-2000,Romero-2000,Yan-2000,stefanseries,dmz-thesis}.
Sabir, \hbox{Bhuiyan}, and Outhwaite \cite{sabir} used the mean spherical approximation (MSA) and two different Poisson-Boltzmann approaches to compute critical parameters resulting from both size and charge asymmetry; 
the reduced critical density, $\rhostar_c$, was always reported to increase with greater size asymmetry, but the trends for the reduced critical temperature, $\tstar_c$, were inconsistent.
(For appropriate definitions of reduced temperature and density, see \cite{Romero-2000} and below.)
Gonzalez-Tovar used the MSA and found, via the energy route, that \emph{both} $\tstar_c$ and $\rhostar_c$ increased with greater size asymmetry \cite{Gonzalez-1999}.
Most recently, Raineri, Routh, and Stell also employed the MSA, but augmented the analysis by incorporating Bjerrum-Ebeling-Grigo pairing; 
they likewise predicted that both $\tstar_c$ and $\rhostar_c$ increase monotonically with size asymmetry \cite{Raineri-2000}.
Recent simulations by Romero-Enrique \emph{et al.} \cite{Romero-2000} (see also \cite{Yan-2000}), however, reveal the \emph{opposite} trends --- namely, that critical temperature and density \emph{decrease} strongly with increasing size asymmetry.
Hence, as remarked in \cite{Romero-2000}, our current state of even qualitative theoretical understanding appears less than adequate.

The present report therefore formulates and analyzes extensions of the \dhtext (DH) approach \cite{debye} to asymmetric primitive models \cite{dmz-thesis}.
We may recall that the original DH theory, when supplemented by Bjerrum ion pairing \cite{bjerrum} and dipolar-pair solvation in the ionic fluid \cite{Fisher-Levin}, remains the most quantitatively successful theory of coexistence and criticality in the size-\emph{symmetric} RPM (see, e.g., [7(b),10]).
%%% above refs not auto-labelled!!!
One might, thus, reasonably hope that a suitable extension of the DH and (DHBjDI \cite{Fisher-Levin}) theories to unequal ion sizes will, at least, yield correct \emph{trends} in the dependence of $\tstar_c$ and $\rhostar_c$ on asymmetry.

In fact, in their original 1923 paper, Debye and H\"{u}ckel \cite{debye} already claimed to have explicit results for asymmetric primitive models (see \cite{dmz-thesis}).
However, serious flaws in this historic formulation suggest other, more systematic extensions of symmetric DH theory appropriate for the asymmetric case \cite{dmz-thesis}.
We analyze these improved ``asymmetric Debye-H\"{u}ckel'' (ADH) theories by comparison with exact series expansions and newly developed bounds, and also assess their general physical character.
We find, in fact, that, contrary to Refs.\ \cite{sabir,Gonzalez-1999,Raineri-2000}, the critical parameters predicted by our modified ADH theories \cite{dmz-thesis} exhibit trends in agreement with those obtained in the recent simulations \cite{Romero-2000,Yan-2000}. 

In order to focus on the effects of size asymmetry, we confine the present discussion primarily to the two-species size-asymmetric primitive model (or SAPM), which consists of equal numbers, $N_+ = N_-$, of positive and negative ions with hard-core diameters $\asubmm \geq \asubpp$, and charges of equal magnitudes $q_+ = -q_- = z_\pm \elcharge$.
(Of course, the complementary case $\asubpp > \asubmm$ follows trivially by symmetry.)
All material and space is assumed to possess a uniform dielectric constant, $D$.
For the most part, we also assume additivity of diameters, 
\begin{equation}
\label{add-diam}
\asubpm \equiv a = \mbox{$\frac{1}{2}$}(\asubpp + \asubmm).
\end{equation}
The degree of asymmetry will be described either by the fractional deviations from $a$, namely, 
\begin{equation}
\label{delta}
\delta_\sigma = (a_{\sigma\sigma} - a)/a \mbox{\hspace{3em}} (\sigma = +, -) \; ,
\end{equation}
or by the diameter ratio \cite{Romero-2000},
\begin{equation}
\label{deltalambda}
\lambda = \asubmm/\asubpp = (1+\delm)/(1-\delm),
\end{equation}
where we have used the fact that $\delp = -\delm$ for the additive case.
Note that the general asymmetric primitive model electrolyte possesses a large --- in fact, three-dimensional --- parameter space: two degrees of freedom for size asymmetry (since additivity need {\em not} be assumed) and one for charge asymmetry (noting that the overall magnitude is irrelevant).

We take the total number density to be $\rho = N/V$, with $V$ the volume and $N = N_+ + N_-$.
The reduced density is then appropriately defined by \cite{Romero-2000}
\begin{equation}
\label{rhostar}
\rhostar = (\rho_+ + \rho_-) a^3 \;,
\end{equation}
which seems the most relevant parameter since the critical densities are typically low, $\rhostar_c \lesssim 0.1$.
But we note that other workers have used alternative definitions \cite{sabir,Raineri-2000}.
The natural temperature scale for studying criticality (i.e., low temperatures) in the SAPM is set by the energy of an {\em attractive} pair of ions at contact, so we take
\begin{equation}
\label{tstargen}
\tstar = k_BT/( |q_+ q_-|/Da_{+-}) \; .
\end{equation}
It is also convenient to define the reduced electrostatic energy, normalized by the energy of closest approach of a $(+,-)$ pair, namely,
\begin{equation}
\label{ue2pm}
u(\rho,T) \equiv U^{\mathrm{ex}}_N(\rho,T)/N(|q_+q_-|/Da),
\end{equation}
where 
$U^{\mathrm{ex}}_N \;\, (\equiv U_N - \threehalves N k_B T)$
is the overall excess energy.

\begin{figure}
\vspace*{1cm}
\centerline{ \epsfxsize = 5in
%\epsfclipon
\epsffile{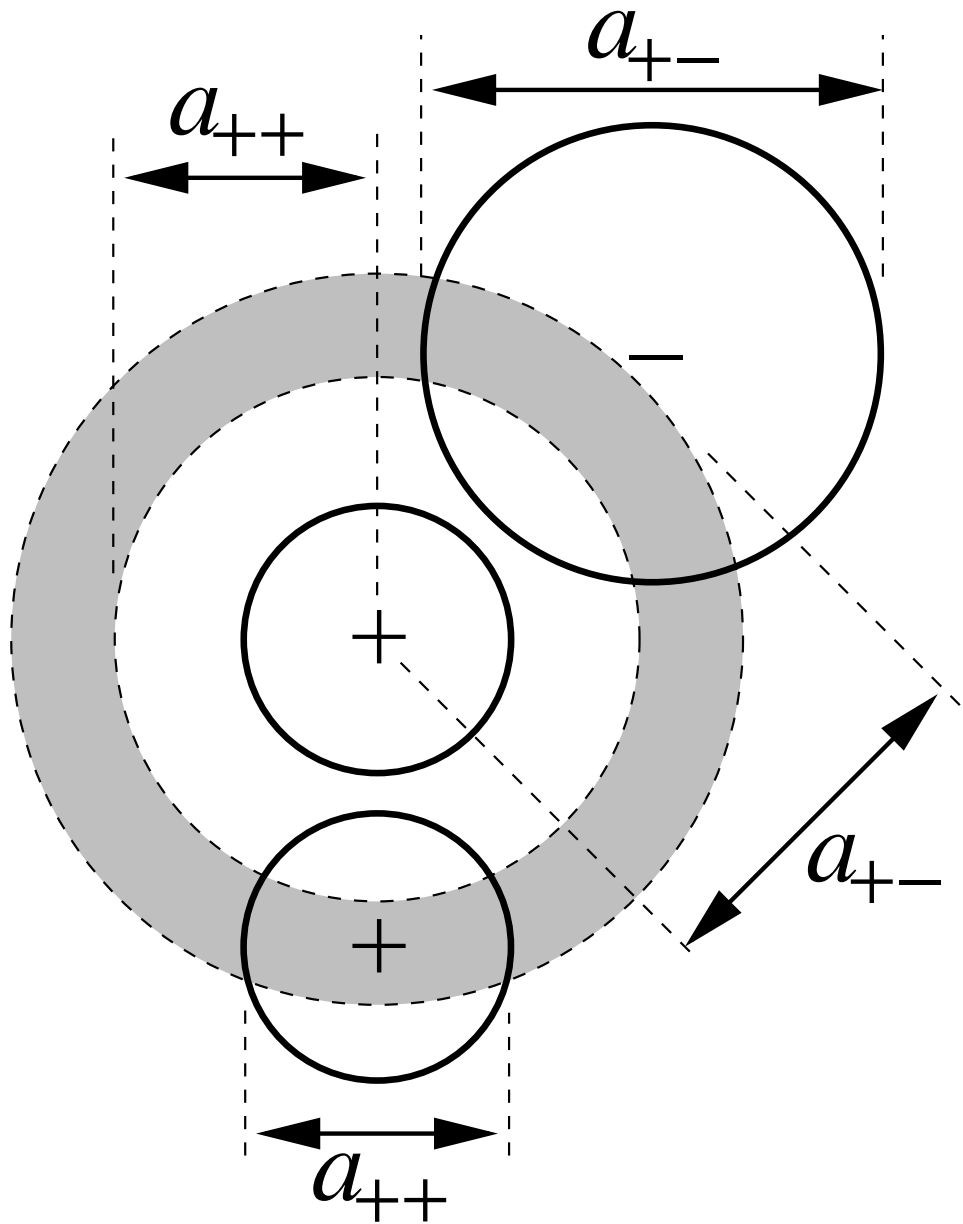}
}
\vspace*{3cm}
\caption
{\label{fig:e2pmlessthan} 
Zuckerman, Fisher, and Bekiranov
}
\end{figure}

The theoretical description of size-asymmetric models is, naturally, more complicated than in the symmetric case.
Specifically, in considering the correlations and fluctuations in the neighborhood of a particular ion (which, following DH, we may suppose is fixed at the origin), \emph{three} distinct surrounding spherical shells or zones must be accounted for.
To see this, suppose first that the selected, fixed ion is positive and that the diameters are ordered in size according to $\asubpp<\asubpm<\asubmm$:
we term this the ``inner'' case since the like-like diameter of the central charge is smaller than $\asubpm$. 
Trivially, no ion center can enter the ``interior'' zone $0 < r < \asubpp$, where $r$ is the radial distance from the origin;
of course, this applies to the symmetric RPM as well.
New to the SAPM, however, is the ``border'' zone (or, more precisely, the ``{\em sub}-border'' zone) $\asubpp < r < \asubpm$ for this ``inner'' scenario, that is shown shaded in Fig.\ \ref{fig:e2pmlessthan}:
this can be populated only by the centers of the smaller, i.e., positive, ions;
the larger negative ions are excluded.
Finally, as in the symmetric model, positive and negative ions may be present in the exterior zone $r > \asubpm$.
When a larger negative ion is chosen to be at the origin, there is a complementary ``super-border'' zone, $\asubpm < r < \asubmm$, into which only positive charges can penetrate.

The presence of these ``charge-unbalanced'' zones in the asymmetric case turns out to play a crucial role which, in particular, means that the simple extension of the DH treatment (ADH theory), although in agreement in leading orders at high $T$ with the exact series expansions \cite{falkeb,stefanseries,dmz-thesis}, fails badly at lower temperatures.
This is shown below in Sec.\ II after we develop the ADH formulation, which allows for the border zones.
While the failure is not entirely surprising, since the general DH formulation relies on a high-temperature expansion, it does present a sharp contrast to the symmetric case. 
One striking consequence of the difficulties is the theory's violation, at low temperatures, of an extension of Onsager's lower bound for the (internal) energy of the RPM \cite{onsag} that we establish in Appendix A for \emph{general} primitive models that are asymmetric in both charge and in size.

It is an interesting fact that analysis of the general primitive model with \emph{nonadditive} diameters presents few additional problems in most cases.
(See, e.g., \cite{stefanseries} and references therein.)
On the other hand, the nonadditive models are potentially useful in applications because they can represent \emph{short range} interactions beyond the pure Coulombic couplings.
Thus if, for example, $\asubpp = \asubmm = a_0$, so that one has a size-\emph{symmetric} model, but $\asubpm > a_0$, a moment's consideration shows that unlike ions experience a strong short-range \emph{repulsion} relative to the geometric ion size $a_0$.
This competes with the ionic attractions and, thus, for example, increases the relative size of a Bjerrum pair and decreases its stability.
Conversely, the case $\asubpm < a_0$ can be viewed as representing an enhanced short range repulsion between \emph{like} ions of a ``true'' diameter less than $a_0$.
In reality, of course, the interactions in ionic systems deviate from pure electrostatics at short distances.
Accordingly, although we mainly focus below on the additive case, we briefly indicate where necessary how to construct DH theories for the general, nonadditive model \cite{dmz-thesis}.
%Accordingly, although we mainly focus below on the additive case, we have recorded in Appendix B a few further details of the DH analysis for the general, nonadditive model \cite{dmz-thesis}.
We believe this theory is of interest in its own right as well as providing a basis for more elaborate treatments.
Further details of the linear asymmetric DH theory are also given in \cite{dmz-thesis}.

Because of the failures of the linear asymmetric DH theory, we have explored some modifications that, for modest degrees of asymmetry, ensure satisfaction of the energy bound of Appendix A and of thermal convexity requirements \cite{critique};
these criteria are also satisfied for large asymmetry in the critical and coexistence regions.
The modified ADH theories accomplish this by specifically allowing for, and compensating in a natural way, the charge imbalances induced by the existence of the border zones \cite{dmz-thesis}.
These theories are expounded in Sec.\ III and their predictions for criticality are examined in Sec.\ IV.
It transpires, as mentioned above, that the simplest modified theories appear to be the first theoretical treatments to predict correctly (as judged by the simulations \cite{Romero-2000,Yan-2000}) that both $\tstar_c$ and $\rhostar_c$ \emph{fall} when size asymmetry is introduced;
however, for $\lambda = \asubmm/\asubpp \gtrsim 4$, the critical density displays incorrect, nonmonotonic behavior.
Finally, Sec.\ V presents a brief overview and some conclusions.

\newpage
\section
{\label{sec:e2pmdhenergy}Asymmetric Debye-H\"{u}ckel Theory}
\label{sec:ladh}

\subsection{Formulation of the theory}
In this subsection we present the ADH theory.
Since the cost in additional complexity is slight, we consider here (and in Appendix A) a general two-species model with point charges $q_\sigma = \zsig \elcharge$ ($\sigma = +,-$) centered in hard spheres of diameters $a_{\sigma\sigma}$ with the collision diameters, $a_{\sigma\tau}$, restricted only by
$\asubpp \leq a \equiv \asubpm \leq \asubmm$.
The \dhtext procedure \cite{debye}, interpreted generally, directs one to calculate the excess thermodynamic functions, based on approximations for the correlation functions which are then integrated via the  ``energy route'' \cite{mcquarrie}.
Specifically, defining $\psi_\sigma$ to be the average electrostatic potential at a (fixed) ion of species $\sigma$ due to all other ions, the reduced electrostatic energy, defined in (\ref{ue2pm}), is given by
\begin{equation}
\label{adhenergy}
u = \frac{Da}{2(\zp-\zm)q_0} \lpar \psi^<_+ - \psi^>_- \rpar \; ,
\end{equation}
where the superscripts $<$ and $>$ merely serve as reminders of the relative ion sizes, and also facilitate formulation of the nonadditive case.
Other thermodynamic quantities, such as the free energy and pressure, follow in principle  after suitable integrations and differentiations.
We derive an explicit closed form expression for the internal energy as a function of $T$ and $\rho$, and the second virial term for the pressure is also discussed.
%For the SAPM, we derive an expression in closed form only for the internal energy;
%however, the second virial term for the pressure is also discussed.

We begin the calculation of the ion potentials $\psi_\sigma(T,\rho)$ by fixing an ion of species $\sigma$ at the origin, as in Fig.\ \ref{fig:e2pmlessthan}.
The induced electrostatic potential, $\phi(\rbf)$, and corresponding charge density, $\rho_q(\rbf)$, are then related by the (exact) {\em averaged} Poisson equation, namely, 
\begin{equation}
\label{poissonavg}
\nabla^2 \langle \phi(\rbf) \rangle_\sigma 
  = -\frac{4 \pi}{D} \langle \rho_q(\rbf) \rangle_\sigma,
\end{equation}
where the subscript $\sigma$ indicates that the averages are performed with a charge of species $\sigma$ at the origin \cite{mcquarrie}.
The ion potentials follow from the limit
\begin{equation}
\label{psifromphi}
\psi_\sigma(T,\rho) = \lim_{r \app 0} \left[ \langle \phi(r) \rangle_\sigma 
		- \frac{q_\sigma}{Dr} \right] \; ,
\end{equation}
which eliminates the self-interaction of the fixed charge at the origin.

The DH procedure then introduces two approximations for the average charge density, $\langle \rho_q(\rbf) \rangle_\sigma$.
The well-known Poisson-Boltzmann approximation is followed by a linearization of the exponential in the Boltzmann factors, yielding 
\begin{eqnarray}
\label{poissonboltz}
\langle \rho_q(r) \rangle_\sigma
  & \simeq & \sum_{\tau = +,-} q_\tau \rho_\tau 
	\exp{\lbrak -\beta q_\tau \langle \phi(r) \rangle_\sigma \rbrak} \; , \\
\label{linearize}
  & \simeq & \sum_{\tau = +,-} q_\tau \rho_\tau
	\lbrak 1 - \beta q_\tau \langle \phi(r) \rangle_\sigma \rbrak \; ,
\end{eqnarray}
where $\beta = 1/k_B T$.
In the size-asymmetric models, the approximate charge density must be allowed to take a different form in each of the three distinct zones around the central charge. 
Thus, when the fixed charge is positive ($\sigma = +$) one has 
\begin{eqnarray}
\label{rhoqnear}
\langle \rho_q(r) \rangle_+ 
  & = & q_+ \delta(r) 
    \mbox{\hspace{4cm}}
    \mbox{for } r < a_{++} \; , \\
\label{rhoqmid}
  & \simeq & q_+ \rho_+ \lbrak 1 - \beta q_+ \langle \phi(r) \rangle_+ \rbrak
    \mbox{\hspace{1.1cm}}
    \mbox{for } a_{++} < r < a \; , \\
\label{rhoqfar}
  & \simeq & -(\kapd^2/4 \pi) \langle \phi(r) \rangle_+ 
    \mbox{\hspace{2cm}}
    \mbox{for } r > a \; ,
\end{eqnarray}
%\begin{equation}
%\label{rhoq}
%\begin{array}{rclcl}
%\langle \rho_q(r) \rangle_+ 
  %& \simeq & q_+ \delta(r) 
    %& \mbox{\hspace{1em}}
    %& \mbox{for } r < a_{++} \; , \\
  %& \simeq & q_+ \rho_+ \lbrak 1 - \beta q_+ \langle \phi(r) \rangle_+ \rbrak
    %& \mbox{\hspace{1em}}
    %& \mbox{for } a_{++} < r < a \; , \\
  %& \simeq & -(\kapd^2/4 \pi) \langle \phi(r) \rangle_+ 
    %& \mbox{\hspace{1em}}
    %& \mbox{for } r > a \; ,
%\end{array}
%\end{equation}
%\begin{equation}
%\label{rhoq}
%\langle \rho_q(r) \rangle_+ \simeq 
%\left \{
%\begin{array}{cll}
%& q_+ \delta(r) & (r < a_{++}) \\
%& q_+ \rho_+ \lbrak 1 - \beta q_+ \langle \phi(r) \rangle_+ \rbrak
	%% = q_+ \rho_+ - (\kappa_+^2/4\pi) \langle \phi(r) \rangle_+
	%\mbox{\hspace{2em}} 
  %& (a_{++} < r < a) \\
%& -(\kapd^2/4 \pi) \langle \phi(r) \rangle_+ & (r > a) .
%\end{array}
%\right.
%\end{equation}
where, as usual \cite{mcquarrie}, overall electroneutrality has been imposed.
Here, the inverse Debye length is defined in the standard way via
\begin{equation}
\label{e2pmkapd}
(\kapd a)^2 = x_D^2 = (4 \pi a^2 q_0^2 / D k_B T)
                         \sum_\sigma \rho_\sigma \zsig^2 \; . 
\end{equation}
Note: 
(i) there is no approximation in the innermost exclusion zone $(r < a_{++})$; 
(ii) the sub-border zone ($a_{++} < r < a = \asubpm$) can contain none of the larger negative ions so that in (\ref{rhoqmid}), which represents the fundamental extension of the original DH theory, only $q_+$ appears on the right-hand side; and 
(iii) the exterior zone $(r > a)$ follows the standard \dhtext form as in the symmetric RPM \cite{debye,mcquarrie}.

%\begin{equation}
%\label{rhoqnear}
%\langle \rho_q(r) \rangle_+ = \rho_q(r) = q_+ \delta(r) \mbox{\hspace{3em}} (r < a_{++});
%\end{equation}
%\begin{equation}
%\label{rhoqmid}
%\langle \rho_q(r) \rangle_+ \simeq 
%  q_+ \rho_+ \lbrak 1 - \beta q_+ \langle \phi(r) \rangle_+ \rbrak
%	= q_+ \rho_+ - (\kappa_+^2/4\pi) \langle \phi(r) \rangle_+
%  \mbox{\hspace{3em}} (a_{++} < r < a);
%\end{equation}
%\begin{equation}
%\label{rhoqfar}
%\langle \rho_q(r) \rangle_+ \simeq 
%   -(\kapd^2/4 \pi) \langle \phi(r) \rangle_+
%  \mbox{\hspace{3em}} (r > a) .
%\end{equation}
%the ``sub-border'' zone, where $+$ but {\em not} $-$ charges may be present  so that, accepting (\ref{rhoqfromg}) - (\ref{linearize}),
%and the ``exterior'' zone, where both charge species are present so that, as in standard DH theory \cite{mcquarrie},

The approximations (\ref{rhoqnear})-(\ref{rhoqfar}) complete the reduction of the averaged Poisson equation (\ref{poissonavg}) to the basic ADH equation for a $+$ ion.
This may be solved for four unknown coefficients using standard electrostatic boundary conditions (continuity of $\phi$ and $\dee \phi / \dee r$) at $r=\asubpp$ and $r=\asubpm=a$ with $\phi(r) \app 0$ as $r \app \infty$ \cite{mcquarrie}, from which $\psi_+$ follows via (\ref{psifromphi}). 
The closed forms for the electrostatic energy in the ADH approximation then result from combining the general expression for the energy (\ref{adhenergy}) with the appropriate expressions for the $\psi_\sigma$ potentials, which are presented below.
If we introduce
\begin{equation}
\thsig \equiv \lbrak \, |\zsig| / (\zp-\zm) \, \rbrak^{1/2}, 
\end{equation}
one finds that the potential in the ``inner'' scenario, with a smaller central positive ion (indicated by the superscript $<$), may be written
\begin{equation}
\label{psiless}
\psi^<_+ \lpar q_0;T,\rho;\{\zsig\},\{\delsig\} \rpar 
   = \lpar \frac{z_+ q_0}{Da} \rpar   
  \frac{ |z_-/z_+| \, \tstar \, ( A^<_1 + A^<_2 + A^<_3 ) \, 
	- \, B^< }{ C^< } \, ,
\end{equation}
where the various contributions are 
\begin{eqnarray}
\label{psilesspart1}
A^<_1 & = & (1+\xd+\xd\delta_+) \lbrak \cosh{(\thp \xd \delta_+)} - 1 \rbrak ,
  \\
A^<_2 & = & - (\thp^{-1}+\thp\xd+\thp\xd\delta_+) \sinh{(\thp \xd \delta_+)} ,
  \\
A^<_3 & = & \xd\delta_+, \\
B^< & = &  \xd \cosh{(\thp \xd \delta_+)} - \thp\xd\sinh{(\thp \xd \delta_+)},
  \\
\label{psilesspart5}
C^< & = & \lpar 1+\xd+\xd\delta_+ \rpar \cosh{(\thp \xd \delta_+)}
	- \thp ( \thp^{-2}+\xd+\xd\delta_+ ) \sinh{(\thp \xd \delta_+)}.
\end{eqnarray} 
Note that $\delta_+ \leq 0$ for this $<$ case.
The $A^<_n$ terms of $\psi^<$ have been grouped so that each vanishes individually when $\delta_+ \app 0$;
in this limit of size symmetry, furthermore, the potential reduces to the standard DH result \cite{debye,falkeb,mcquarrie} for a size-symmetric primitive model, namely,
$\psi_+^< = \psi_+ = -(z_+ q_0/Da) \xd/(1+\xd)$.
When the diameter of the negative ion is less than $a$, one should simply switch the species subscripts in (\ref{psiless}).

For the ``outer'' situation, with a larger central ion (indicated by superscript $>$), the corresponding result is
\begin{equation}
\label{psigrtr}
\psi^>_-(q_0;T,\rho;\{\zsig\},\{\delsig\}) = \lpar \frac{|z_-| q_0}{Da} \rpar   
  \frac{ \tstar ( A^>_1 + A^>_2 + A^>_3 ) \, 
	+ \, B^> }{ C^> } \, ,
\end{equation}
where the contributions are now 
\begin{eqnarray}
\label{psigrtrpart1}
A^>_1 & = & (1+\xd) \lbrak \cosh{(\thp \xd \delta_-)} - 1 \rbrak ,
  \\
A^>_2 & = & (\thp^{-1}+\thp\xd) \sinh{(\thp \xd \delta_-)} ,
  \\
A^>_3 & = & -\xd\delta_-, \\
B^> & = &  \xd \cosh{(\thp \xd \delta_-)} + \thp\xd\sinh{(\thp \xd \delta_-)},
  \\
\label{psigrtrpart5}
C^> & = & \lpar 1+\xd \rpar \cosh{(\thp \xd \delta_-)}
	+ ( \thp^{-1}+\thp\xd ) \sinh{(\thp \xd \delta_-)}.
\end{eqnarray} 
Naturally, $\psi^>_-$ also reduces to the symmetric DH result when $\delta_- \app 0$.
(If $\asubpp > \asubmm$, one should simply switch the $+$ and $-$ subscripts.)

For future reference, notice that both $\psi_+^<$ and $\psi_-^>$ have readily-calculated zero-temperature limits, namely,
\begin{equation}
\label{psiatzerot}
\psi^<_\sigma(T \! = \! 0) = - \lpar \frac{z_\sigma q_0}{Da} \rpar \frac{1}{1+\delta_\sigma}
\mbox{\hspace{2em} and \hspace{2em}}
\psi^>_\sigma(T \! = \! 0) = - \frac{z_\sigma q_0}{Da} \; ,
\end{equation}
which are independent of density.
Furthermore, small diameters, $a$, are of interest; but $\psi^<_\sigma$ {\em diverges} for ``point'' ions, i.e., in the limit $a_{\sigma\sigma} \app 0$ (or $\delta_\sigma \app -1$).
This, in fact, serves as a warning sign, as will be seen below.

\subsection{Assessment of ADH Theory}
We now examine the predictions of the linear ADH theory.
Following Debye and H\"{u}ckel \cite{debye,mcquarrie}, the free energy is to be obtained from the internal energy in (\ref{adhenergy}) by employing the Debye charging process (which is equivalent to using the standard thermodynamic relation for the free energy in terms of the energy).
This yields the reduced excess free energy density as
\begin{eqnarray}
\label{dhcharging}
\fbar^{\mathrm ex} (T,\rho;\{\zsig\},\{\delsig\})
  & \equiv & -A_N^{\mathrm{ex}} (V; T) / V k_B T \; , \nonumber \\
  & =      & \frac{-D \kapd^2}{4 \pi q_0 (\zp - \zm)}
	\int_0^1 {\mathrm d} \zeta \, 
	\lbrak \psi_+^<(\zeta q_0) - \psi_-^>(\zeta q_0) \rbrak,
\end{eqnarray}
where $A_N^{\mathrm{ex}}$ is the total excess Helmholtz free energy.
It must be recalled that in addition to the explicit dependence on $q_0$ (and, hence, on $\zeta$) entering (\ref{adhenergy}) via (\ref{psiless}) and (\ref{psigrtr}), an implicit dependence occurs via $\kapd \propto q_0$ which enters (\ref{psilesspart1})-(\ref{psilesspart5}) and (\ref{psigrtrpart1})-(\ref{psigrtrpart5}).

Except in the standard symmetric case ($\lambda = 0$, $\delp = \delm = 0$) most of the integrals involved in (\ref{dhcharging}) seem intractable.
Nevertheless, one may derive an expansion for the free energy, and thence for all other thermodynamic quantities, such as the pressure, by expanding the expressions (\ref{psiless})-(\ref{psigrtrpart5}) and integrating term by term.
By this route, we can check the theory against exactly known expansions (see, e.g., \cite{falkeb,stefanseries,dmz-thesis}).

Thus, consider the low-density/high-temperature expansion for the general primitive model (with arbitrary charges $q_\sigma = z_\sigma q_0$ and diameters $a_{\sigma \tau}$) which is known to overall order $\rho^{5/2}$ in the density.
On specializing to the two-component case, this may be written
\begin{eqnarray}
\label{fbarexact}
\fbar(T,\rho) 
  & = & \kapd^3 / 12 \pi
        + B^\dagger_{2,3} \rho^2 \ln{(\kapd a)} / {\tstar}^3
        \nonumber \\
  &   & \mbox{} + \rho^2 \sum_{n=0}^\infty B_{2,n}  / {\tstar}^n
        + O[ \rho^{5/2} \ln{(\kapd a)} ] \; ,
\end{eqnarray}
where we again recall $\kapd a \equiv x_D \propto \sqrt{\rhostar}$.
The leading term here varies as $\rho^{3/2}$ and represents the DH limiting law reproduced by all sensible approximations.
The coefficient $B_{2,0}$ derives from the second virial coefficient for a pure hard-sphere gas.
At the present level of approximation, this may be accounted for precisely by including the hard-core free energy density, $\fbar^{\mathrm{HC}}$, in the basic approximation \cite{Fisher-Levin}.
The remaining coefficients
$B_{2,1}, B_{2,2}, B^\dagger_{2,3}, B_{2,3}, B_{2,4}, \ldots$
arise from the electrostatic interactions.

Now, DH-based approximations (in common with the MSA, etc.) cannot generate the $\rho^2 (\ln{\rho})/T^3$ term with coefficient $B^\dagger_{2,3}$.
However, our linear ADH theory yields the exact leading coefficients in order $1/T$ and $1/T^2$.
Explicitly we find
\begin{equation}
\label{b21}
B_{2,1} = \pi a^3 \, |\zp \zm| \lbrak \frac{\delp(2+\delp) + \delm(2+\delm)}{(\zp - \zm)^2} \rbrak \; ,
\end{equation}
which vanishes for the size-symmetric models (with $\delsig = 0$), and
\begin{equation}
\label{b22}
B_{2,2} = -\pi a^3 \lbrak 1 + \frac{\zp^2 \delp + \zm^2 \delm}{(\zp - \zm)^2} \rbrak \; .
\end{equation}
%Note that in writing these expressions for the general two-species primitive model with charge asymmetry allowed, we have used some of the further details given in Appendix B.
It is, indeed, striking that the exact leading order dependences in powers of $1/T$ are reproduced for arbitrary diameters, $\asubpp$, $\asubpm$, and $\asubmm$ (including nonadditivity and point charges).
However, one probably should not be so surprised since the arguments leading to the basic ADH equations, (\ref{poissonavg}) with (\ref{poissonboltz}) and (\ref{linearize}) are, on reflection, clearly valid to leading orders in $\rho$ and $1/T$ even if it is not obvious how far the validity will go. 

\begin{figure}
\centerline{ \epsfxsize = 4.5in
%\epsfclipon
\epsffile{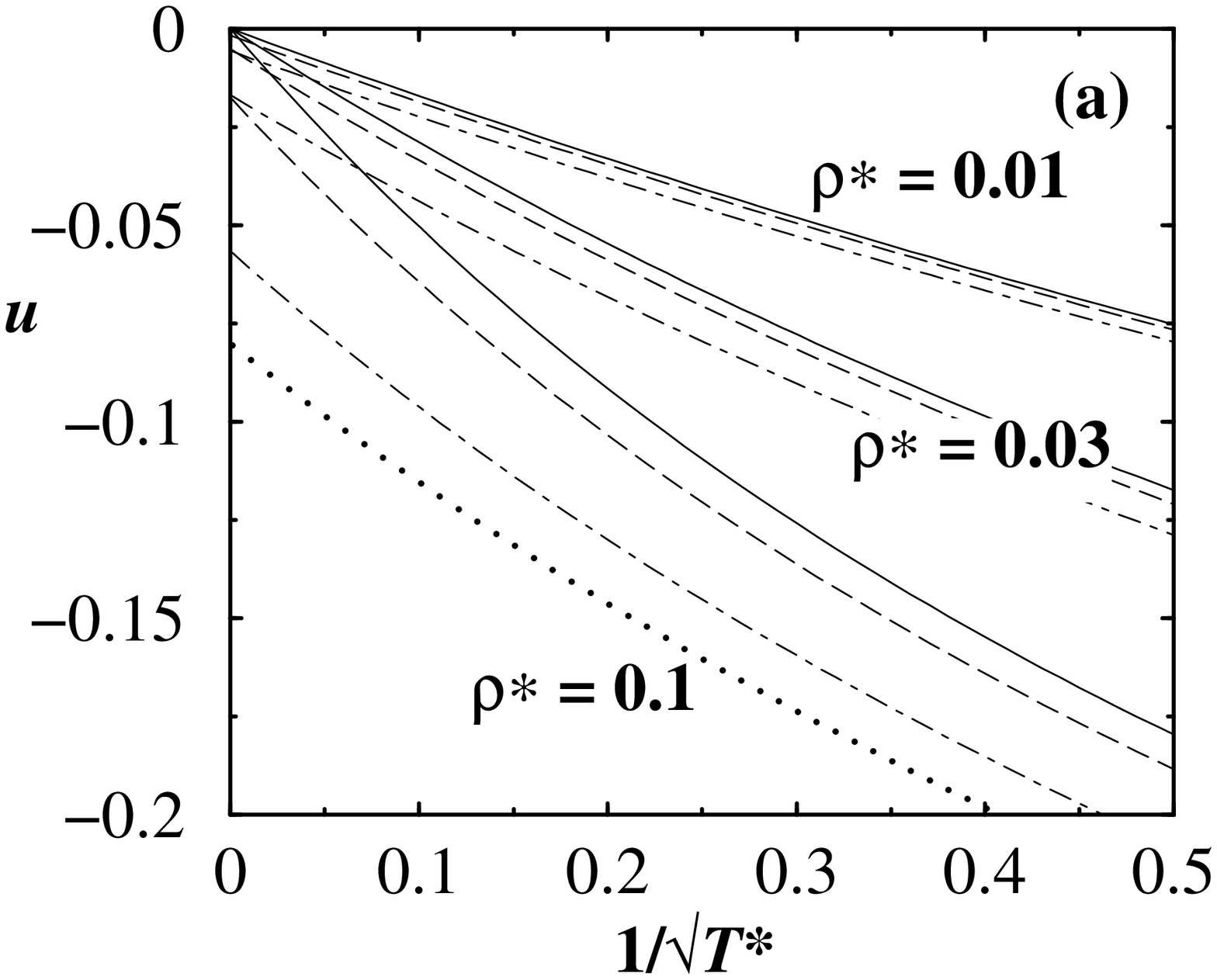}
}
\centerline{ \epsfxsize = 4.5in
%\epsfclipon
\epsffile{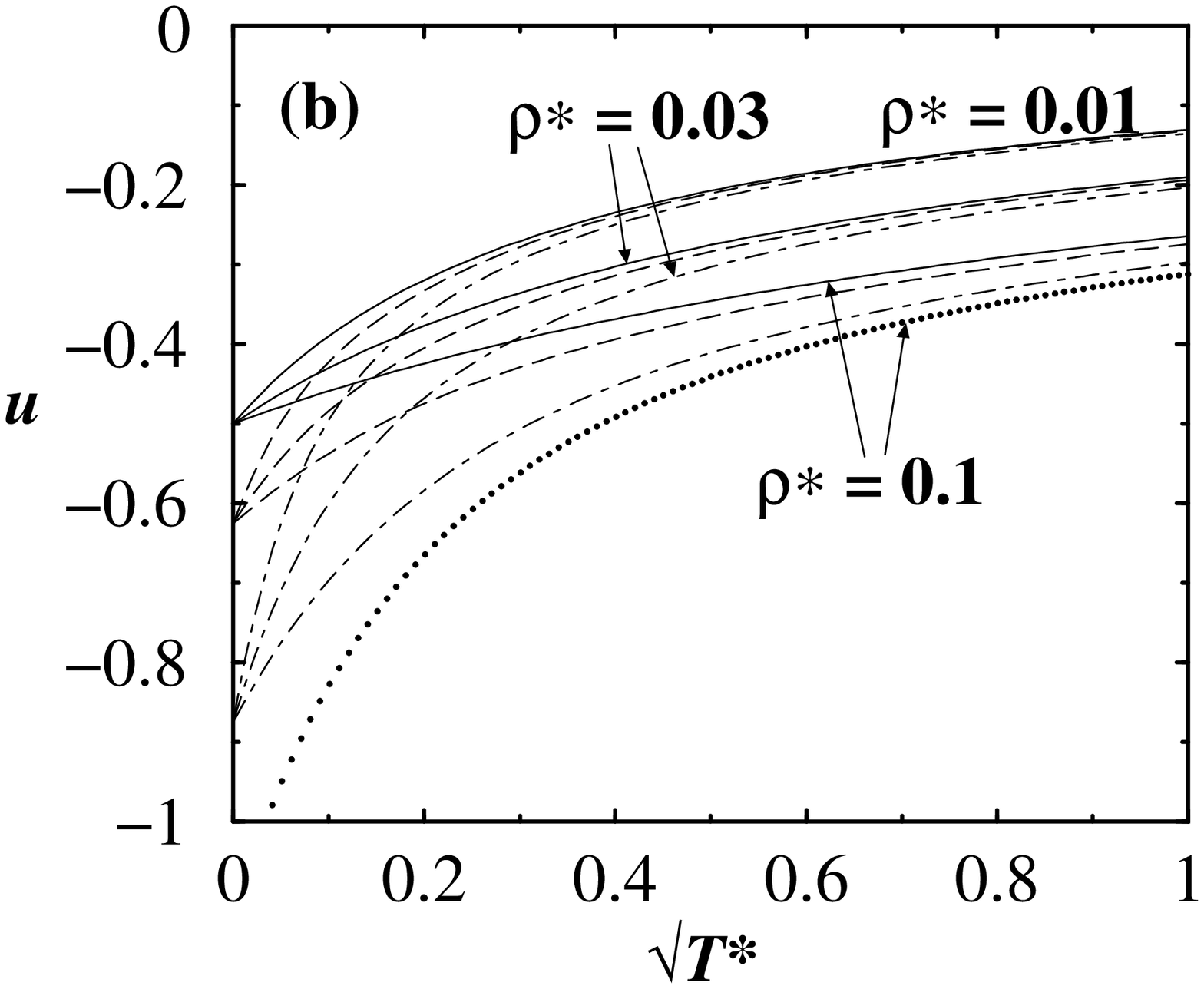}
}
\vspace*{2cm}
\caption
{\label{fig:uladh}
Zuckerman, Fisher, and Bekiranov
}
\end{figure}

To study the ADH theory more quantitatively, we focus on the energy for the additive 1:1 case ($\zp = - \zm = 1$).
No integrations are then required.
Fig.\ \ref{fig:uladh} displays energy isochores as functions of $\tstar$ at high and low temperatures for degrees of asymmetry
$\lambda = \asubmm/\asubpp = 1$ (the RPM for reference, solid curves), 2, 4 and 6.
These values of $\lambda$ correspond to size deviations
$|\delp| = \delm = 0, \mbox{$\frac{1}{3}$}, \mbox{$\frac{3}{5}$}$ and $\mbox{$\frac{5}{7}$}$, respectively.
It is evident that ADH theory predicts that size-asymmetry lowers the electrostatic energy, with the effects being greatest at the highest and lowest temperatures.
At intermediate $T$ the predictions for varying $\lambda$ are grouped together according to density.

Notable features of the high-$T$ plots of Fig.\ \ref{fig:uladh}(a) are, first, that the finite values of $u$ attained when $T \app \infty$ are exact.
This can be verified by a simple \emph{a priori} calculation;
alternatively, the values may be checked from the exact order $\rho^2/T$ term as displayed in (\ref{fbarexact}) and (\ref{b21}) that is correctly reproduced by ADH theory.
The limiting slopes at $\tstar = \infty$ are discussed in further in \cite{dmz-thesis}.

The behavior of $u(T,\rho)$ at low temperatures proves, however, much less satisfactory.
From Fig.\ \ref{fig:uladh}(b) one sees that the energy is predicted to fall increasingly rapidly for increasing asymmetry when $T$ falls, passing well below the ground-state prediction $u = -\mbox{$\frac{1}{2}$}$ of the symmetric DH theory; 
but see also \cite{critique} for the DHBjDI extensions \cite{Fisher-Levin}.
Most seriously, however, for large enough asymmetry, the energy drops below the 1:1 size-asymmetric lower bound, namely, $u \geq -1$, established in Appendix A (which also treats general $\zp$ and $\zm$).
More concretely, the ground states predicted by ADH theory follow from (\ref{adhenergy}) and (\ref{psiatzerot}) which yield
\begin{equation}
\label{uadhzerot}
u^{\mathrm{ADH}}(T\!=\!0; \lambda) = - \mbox{$\frac{1}{8}$} (3+\lambda) \; .
\end{equation}
This expression evidently violates the bound when $\lambda > 5$ 
( or $|\delp| = \delm > \frac{2}{3}$).

Of course, this is unacceptable, but even for smaller values of $\lambda$ this behavior casts doubt on the value of the approximation at low $T$ in the vicinity of the expected critical region.
The origin of this behavior can be understood physically by examining the ADH predictions for the mean total charge, say $Q^<_+$, within the sub-border zone $\asubpp < r < \asubpm = a$ which can only be penetrated by like, i.e., positive charges.
Recalling the notations in (\ref{poissonavg}) and the approximations (\ref{poissonboltz})-(\ref{rhoqfar}), the theory yields
\begin{eqnarray}
\label{zonecharge}
Q^<_+  
  & =      & \int_{\asubpp}^{\asubpm} \dtrstd  \langle \rho_q(r) \rangle_+ \nonumber \\
  & \simeq & \int_{\asubpp}^{\asubpm} \dtrstd  q_+ \rho_+ \lbrak 1 - \beta q_+ \langle \phi(r) \rangle_+ \rbrak \; ,   
\end{eqnarray}
where the precise form of $\langle \phi(r) \rangle_+$ follows from the explicit solution of (\ref{poissonavg}) with (\ref{rhoqnear})-(\ref{rhoqfar}) \cite{dmz-thesis}.
Now, when $T\app\infty$ the approximation becomes exact, yielding
$Q^<_+ = q_+ \rho_+ V^<$, where 
$V^< = \frac{4}{3} \pi (\asubpm^3 - \asubpp^3)$
is the volume of the sub-border zone (see Fig.\ 1).
At finite but large temperatures, (\ref{zonecharge}) correctly predicts the linear decrease of $Q^<_+$ with $\beta$ as the $(+,+)$ repulsions come into play.
Of course, the repulsions become increasingly effective as $T$ decreases so that, at low $T$, one expects $Q^<_+$ to remain positive but to be of magnitude only
$V^< q_+ \rho_+ \exp{(-\beta q_+^2 / D \asubpm)}$ (although, because, of the negative screening cloud that builds up for $r > \asubpm$, this estimate might need to be modified somewhat).

However, if one neglects the screening, the mean potential $\langle \phi(r) \rangle_+$ in (\ref{zonecharge}) will be of order
$+q_+ / D \asubpm$, or larger, in the sub-border zone.
Then, as $T$ falls and $\beta \app \infty$ it is evident that (\ref{zonecharge}) is likely to predict, first, a totally unphysical \emph{negative} value for $Q^<_+$ (even though \emph{no} negative charges can enter the sub-border zone) and, eventually, a divergence of $Q^<_+$ to $-\infty$.
Explicit calculations \cite{dmz-thesis} fully bear this out.
For example, when $\lambda = 2$, and $\rhostar = 0.01$, one finds negative $Q^<_+$ for $\tstar \lesssim 0.9$;
for $\rhostar = 0.1$ ($\gtrsim \rhostar_c$) the change of sign is delayed until $\tstar \simeq 0.6$ but the divergence to $-\infty$ is more rapid.
Since the ADH theory for the border zones embodied in (\ref{linearize}), (\ref{rhoqmid}), and (\ref{zonecharge}) reflects a \emph{high}-$T$ expansion, 
the problems at low $T$ should not be a great surprise.
On the other hand, the successes of the original DH theory for the symmetric case at low $T$ (e.g., \cite{Fisher-Levin,critique}) clearly hinge on the absence of any ``charge-unbalanced'' border zones.

Clearly the serious physical defects of the ADH treatment must be rectified if the theory is to have any value for $\tstar \lesssim 1$:
and, we expect, $\tstar_c \lesssim 0.1$ \cite{fisher,stellrev,Fisher-Levin,critique,sabir,Gonzalez-1999,Raineri-2000}.
Before addressing that task, however, we present, briefly, the original claims of Debye and H\"{u}ckel \cite{debye} as regards a system of ions with arbitrary diameters.

\subsection{Original DH Theory}
In their original 1923 paper, Debye and H\"{u}ckel \cite{debye} did, in fact, discuss the case of an ``arbitrary ionic solution'' --- that is, within the model they adopted, a mixture of hard-spheres of varying diameters $a_{\sigma\sigma}$ and charges $q_\sigma$, or the \emph{general} primitive model.
However, in their analysis no mention is made of the border zones in which charges of only species with small enough diameters can be present.
(Note that in the multispecies case there will, in general, be a number of \emph{distinct} border zones around each charge.)
Rather, their argument is presented \emph{as if} all species were of the \emph{same} diameter \cite{debye,dmz-thesis}.
The mathematics is then identical to that for the RPM.
In our notation, Debye and H\"{u}ckel thus present the conclusion
\begin{equation}
\label{epmpsiofDandH}
\psi_\sigma = - \lpar \frac{\zsig q_0}{D} \rpar 
	\frac{\kapd}{1 + \kapd a_{\sigma\sigma}},
\end{equation}
(i.e., precisely the RPM result when $a_{\sigma\sigma} \equiv a$).
The equivalent free energy for a 1:1 model is then
\begin{eqnarray}
\label{fbar1923}
a^3 \fbar^{\mathrm{DH}} & = &
\frac{1}{16\pi} \left\{ 
  \frac{ 2 \ln{[1 + (1+\delp)\xd]} + [(1+\delp)\xd]^2 - 2 (1+\delp)\xd}
	{ (1+\delp)^3 }
\right. \nonumber \\
&& \hspace{1cm} \mbox{}+ 
\left.
  \frac{ 2 \ln{[1 + (1+\delm)\xd]} + [(1+\delm)\xd]^2 - 2 (1+\delm)\xd}
	{ (1+\delm)^3 }
\right\} \;,
\end{eqnarray}
from which predictions for criticality etc.\ follow: see below.

Neglecting all the border zones spares this ``original DH'' theory the pathology of our ADH theory, but seriously reduces the plausibility of their treatment for size-asymmetric models.
Thus, not so surprisingly, the high-$T$/low-$\rho$ expansions of the thermodynamics resulting from (\ref{fbar1923}) differ at the first correction to the limiting behavior --- see $B_{2,1}$ in (\ref{b21}) --- from the exact results captured by our ADH theory;
indeed, the former depends only on the ratio $\xd^2 \propto\rho/T$, while the exact results are more complex.

\newpage
\section{\label{sec:e2pmdhvariants} Modified Asymmetric DH Theory}
We now discuss a class of modifications of ADH theory (originally introduced in \cite{dmz-thesis}) which avoid the unphysical behavior of the charge density in the border zones:
this behavior was identified in Sec.\ II.B above as a root cause of the pathological low-$T$ behavior which included violations of the lower bound on the internal energy (Appendix A).
In the following section, we examine the predictions of these modified ADH theories for the critical parameters (including a comparison with the original, 1923 DH proposal for asymmetric systems).

\subsection{Introduction of Border Zone Factors}
It is clear from the previous considerations of the ADH prediction (\ref{zonecharge}) for the charge $Q^<_+$ in a sub-border zone, that the standard DH linearization of the Poisson-Boltzmann exponential is generally inadequate in any zone where the mean charge is necessarily unbalanced because of strongly repulsive short-range, steric repulsions.
As an alternative first step, that will still yield an analytically tractable theory, we forego the self-consistent aspect of the Poisson-Boltzmann approximation (\ref{poissonboltz}) in the border zones and consider replacing the self-consistently determined electrostatic potential
$\langle \phi(r) \rangle_\sigma$ by a \emph{fixed}, but possibly temperature- and density-dependent effective potential, 
$\phi^\dagger_\sigma(r; T, \rho)$.

One possibility is merely to use the \emph{bare} potential, i.e., to put
$\phi^\dagger_\sigma = q_\sigma/Dr$.
The right-hand side of (\ref{rhoqmid}) would then read
$q_+ \rho_+ \exp{(-\beta q_+^2 / Dr)}$;
but an obvious drawback of such an approximation is that the resulting closure of Poisson's equation, (\ref{poissonavg}), is no longer readily solvable.
Instead, we simplify further by neglecting the $r$-dependence of $\phi^\dagger_\sigma$.
Thus we introduce temperature-dependent \emph{border zone factors} $\eless(T)$ and $\egrtr(T)$ and consider the replacement of (\ref{rhoqmid}) for a sub-border zone by
\begin{equation}
\label{rhoqmidmefless}
\langle \rho_q(r) \rangle_+ 
  \equiv \rho_q^<(r;T,\rho) \simeq q_+ \rho_+ \eless(T),
  \mbox{\hspace{3em}} a_{++} < r < a \; .
\end{equation}
Likewise, for a super-border zone (indicated, as before, by the superscript $>$) we advance
\begin{equation}
\label{rhoqmidmefgrtr}
\langle \rho_q(r) \rangle_-
  \equiv \rho_q^>(r;T,\rho) \simeq q_+ \rho_+ \egrtr(T),
  \mbox{\hspace{3em}} a < r < a_{--}.
\end{equation}

The calculation of the $\psi_\sigma$ potentials now proceeds just as in Sec.\ \ref{sec:ladh} but the results take a simpler form, which we distinguish by using a circumflex. 
Maintaining our convention of a smaller positive ion ($\delta_+ < 0$) and larger negative ion ($\delta_- > 0$), we find
\begin{eqnarray}
\label{psilessfisher}
\psihat_+^< & = & -\frac{z_+ \elcharge}{Da} \frac{\xd}{1+\xd}
  \left\{1 + \xd \frac{|\zm| \tstar \delta_+ \eless(T)}{2 (\zp-\zm)}
	\lbrak 2 + \delta_+ - \xd (\delta_+ + \twothirds \delta_+^2) \rbrak
  \right\} \; , \\
\label{psigrtrfisher}
\psihat_-^> & = & -\frac{z_- \elcharge}{Da} \frac{\xd}{1+\xd}
  \left\{1 + \xd \frac{\zp \tstar \delta_- \egrtr(T)}{2 (\zp-\zm)}
	\lbrak 2 + \delta_- + \xd (\delta_- + \boxthird \delta_-^2) \rbrak
  \right\} \; . 
\end{eqnarray}
These modified potentials, valid for general $\egenl(T)$, reproduce the symmetric DH theory when $\delta_\sigma \app 0$, and always generate the standard limiting laws when $\rho \app 0$.
%When $T \app \infty$, we note that if $\cal F$ behaves as a constant, then the corrections to symmetric DH theory (which asymptotically behave as $x_D^2 T \cal F$) will be $T$-independent and alter the simple DH thermodynamic predictions in that limit.
For large $\xd \propto (\rho/T)^{1/2}$, these approximations for the $\psi_\sigma$ behave as $x_D^2 T \cal F$, whereas the standard DH expressions approach constants.
%These facts suggest that some choices for $\cal F$ may produce undesirable features;
While the consequences of this fact are not obvious, some numerical results are discussed in Sec.\ III.C, below.

On the other hand, both of the high-$T$ second-virial terms, $B_{2,1}$ and $B_{2,2}$ that are correctly generated by the ADH theory [see (\ref{b21}) and (\ref{b22})], will be reproduced now if the zone factors satisfy
\begin{eqnarray}
\label{zonefacexpandless}
\eless(T) & = & 1 - \frac{ |\zp/\zm| }{ \tstar (1+\boxhalf \, \delp) } 
  + O \! \lpar \frac{1}{{\tstar}^2} \rpar \; , \\
\label{zonefacexpandgrtr}
\egrtr(T) & = & 1 + \frac{ 1 }{ \tstar (1+\boxhalf \, \delm) } 
  + O \! \lpar \frac{1}{{\tstar}^2} \rpar \; .
\end{eqnarray}
In principle, $\eless$ and $\egrtr$ could be chosen to generate further exact second-virial terms; 
but such an approach does not seem of much practical value.

\subsection{Choice of Zone Factors}
Undoubtedly the simplest reasonable approximation for the zone factors is provided by taking 
\begin{equation}
\label{mfzonefac}
\mbox{MF: \hspace{1cm}}   \eless(T) = \egrtr(T) = 1,
\end{equation}
or, equivalently, by dropping the term linear in $\beta$ in (\ref{linearize}) and (\ref{rhoqmid}).
The results will fail to reproduce the correct $B_{2,2}$ coefficient in (\ref{b22}), although the exact $B_{2,1}$ term will be generated.
Physically, this approximation amounts to a direct mean-field or van-der-Waals-type approach in which the effects on the internal energy of the ($T = \infty$) unbalanced zone charges, 
$Q^<_+ = q_+ \rho_+ V^<$ and 
$Q^>_- = q_- \rho_- V^>$ (where $V^<$ and $V^>$ are the zone volumes), are accounted for in a direct way that neglects fluctuations.
The zone charges remain finite and, in fact, \emph{fixed} for all $T$:
but, at least for small relative zone volumes
$V^</a^3 \simeq 4 \pi \delp (1+\delp)$ and
$V^>/a^3 \simeq 4 \pi \delm (1+\delm)$,
one might reasonably expect that the initial thermodynamic trends with increasing asymmetry $\lambda = \asubmm/\asubpp$, for small $\lambda$, will be correctly predicted. 
%Difficulties do, in fact, arise for very extreme parameters and densities:

A second natural step, following our initial discussion, is to take $\phi^\dagger(r)$ to be the direct potential, $q_\sigma/Dr$, evaluated at the \emph{mean radius} of the border zones, namely,
$r^< = \boxhalf (\asubpp + a) = (1+\boxhalf \delp) a$ and
$r^> = (1+\boxhalf \delm) a$.
This amounts to adopting
\begin{equation}
\label{expzonefac}
\mbox{EXP$^<$: \hspace{1cm}}    \eless(T) = \exp{[-\beta q_+^2/Da(1+\boxhalf \delp)]} \; ,
\end{equation} 
and similarly for $\egrtr(T)$ but with the exponent
\begin{equation}
\label{expogrtr}
y^> = +\beta q_+ q_- / Da (1 + \boxhalf \delm) \; .
\end{equation}
It is encouraging that these choices precisely satisfy (\ref{zonefacexpandless}) and (\ref{zonefacexpandgrtr}) and so reproduce \emph{both} $B_{2,1}$ and $B_{2,2}$.

As regards the sub-border zone where, as discussed in Sec.\ II.B, one expects $Q^<_+(T,\rho)$ to vanish at low $T$ (or, at least, approach very small values), the assignment (\ref{expzonefac}) seems rather satisfactory.
Indeed, combining (\ref{rhoqmidmefless}) and the first part of (\ref{zonecharge}) shows that $Q^<$ will remain positive but decrease exponentially rapidly as the $(+,+)$ repulsions ``flush out'' charge from the sub-border zone.

It is worth pointing out that the exponential treatment of the sub-border zone can be extended within the DH self-consistent spirit by linearizing the Poisson-Boltzmann factor in (\ref{linearize}) about the central value of the direct potential in the sub-border zone.
Thus, if $-y^<$ denotes the exponent in (\ref{expzonefac}), while 
$y = \beta q_+ \langle \phi(r) \rangle_+$,
one may replace (\ref{rhoqmid}) by
\begin{eqnarray}
\label{zonefacexplin}
\langle \rho_q(r) \rangle_+ 
  & \simeq & q_+ \rho_+ e^{-y} = q_+ \rho_+ e^{-y^<} e^{-(y-y^<)} 
  \nonumber \\
  & \simeq & q_+ \rho_+ e^{-y^<} \lbrak 1 - (y-y^<) \rbrak \; .
\end{eqnarray}
The overall factor $e^{-y^<}$ should now ensure the sensible behavior of the sub-border zone charge, $Q^<_+$, while the linear factor $y \propto \langle \rho_q(r) \rangle_+$ accounts self-consistently for some fluctuation effects while still allowing integration of the closed ADH equations.
We have not, however, examined this approach further.

In the super-border zone case, however, matters are significantly different.
Certainly, as follows from (\ref{zonefacexpandgrtr}), one expects the mean border zone charge $Q^>_-(T,\rho)$, which will be positive for $T=\infty$, to increase initially when $T$ falls.
This is simply because the Boltzmann factor enhances the attractions between the larger, central negative ion and the smaller positive ions.
Nevertheless, the growth of $Q^>_-$ cannot continue indefinitely as would be implied by adopting the EXP$^>$ form with (\ref{expogrtr}) for $\egrtr$.
Regardless of other effects, the hard cores of the $+$ ions must limit the number that can be packed into a super-border zone at any $T$:
hence $Q^>_-$ \emph{must saturate}.
This effect may be incorporated into the present framework via a saturation fraction, $\egrtr_\infty = s$, which prescribes the limiting $T \app 0$ super-border zone charge density as $s q_+ \rho_+$; recall (\ref{rhoqmidmefgrtr}).
A plausible zone factor is then the ``regulated'' exponential form
\begin{equation}
\label{regexpzonefac}
\mbox{REGEXP$^>$: \hspace{1cm}}    
\egrtr(T) = \frac{ s \, e^{c y^>} }{ e^{c y^>} - 1 + s } \; ,
\end{equation}
with $c = s/(s-1)$, which satisfies (\ref{zonefacexpandgrtr}) and approaches $s$ for large $y^>$ (i.e., low $T$).
In the limit $s \app 1$ (no growth in $Q^>_-$) this reduces simply to the mean-field form (\ref{mfzonefac}).

The difficulty in utilizing this proposal, however, is to know what saturation value ($s > 0)$ is appropriate.
At low $T$, considerations of ion association indicate that the predicted saturation charge in the super-zone, namely, 
\begin{equation}
\label{supersat}
Q^>_{-s} = V^> q_+ \rho_+ s = 4 \pi q_0 
            \frac{ \delm (1 + \delm + \boxthird \delm^2) }
                 { 1 + |\zp/\zm| }
            \zp \rhostar s \; ,
\end{equation}
should not exceed the neutralizing value $|q_-| = |\zm| q_0$. 
To meet this condition requires $s \propto 1/\rho$.
While a density-dependent $s$ and, hence, $\egrtr$, can be accommodated without changing (\ref{psigrtrfisher}), the subsequent expressions for the free energy become more complex.
Furthermore, when $\rhostar$ increases, $s$ must decrease and may even fall \emph{below} unity.
In these circumstances and in the absence of other effective selection criteria we believe it is appropriate to accept the mean-field form (\ref{mfzonefac}), i.e., to set $\egrtr \equiv 1$ even though $B_{2,2}$ will not then be correctly reproduced.

In passing we mention, nonetheless, that we have also explored \emph{algebraic} forms which respect (\ref{zonefacexpandgrtr}) \cite{mef-dmz-unpub}.
As an example, one can consider
\begin{equation}
\label{zonefacalg}
\egrtr(T) = (1 + y^>) / [ 1 + (y^>/s)^2 ]^\nu \; ,
\end{equation}
which saturates at $\egrtr = s$ (not necessarily greater than unity) when $\nu = \boxhalf$ but decays to 0 at low $T$ when $\nu > \boxhalf$.
For $s \gtrsim 2$ such approximations have an added, unexpected feature, namely, they tend to predict an additional asymmetric critical point at higher densities that does not smoothly connect to the standard DH critical point \cite{Fisher-Levin} when $\lambda \app 0$.
While, owing to the strong short-range interactions implicit in the asymmetric systems, such extra critical points are not obviously unphysical (especially when the diameters are nonadditive), we will not discuss the algebraic forms (\ref{zonefacalg}) (or EXP$^>$) further here.

\subsection{Behavior of Energy Isochores}
Before examining the predictions of the modified ADH theories for criticality, it is instructive to examine energy isochores, such as those plotted in Fig.\ 3, for several densities near the critical value:
compare with the results for the simple ADH theory in Fig.\ 2(b) but note the difference in vertical scales.
At temperatures in the coexistence region ($\tstar \lesssim 0.2$) and for modest degrees of asymmetry ($\lambda \lesssim 3\mbox{ - }4$), both the MF-ADH theory [specified by (\ref{mfzonefac})] and the EXP$^<$MF$^>$ADH theory [where $\eless$ has been replaced by (\ref{expzonefac})] predict that the internal energy \emph{increases} above the RPM value with increasing asymmetry.
This reduction in thermodynamic stability suggests, as we will confirm, that the predicted values of $\tstar_c$ fall as $\lambda$ increases. 

We also find that the MF-ADH and EXP$^<$MF$^>$ADH theories do, indeed, satisfy the energy bounds of Appendix A for all $(\rho,T)$ states relevant to the critical and coexistence regions in the additive case.
Bound violations can occur, but these arise only at the highest densities ($\rhostar \simeq 1$) and when the asymmetry is great ($\lambda \gg 1$).
While this behavior undermines the two modified ADH theories investigated as overall descriptions of the asymmetric primitive model, the pathologies occur far from the critical region and at unphysically large asymmetry levels.

For moderate to large asymmetry and high density, however, we find that the MF-ADH and \expmf isochores exhibit \emph{nonmonotonic} behavior in $T$, indicating a thermodynamic instability as previously found in a variety of ion-pairing theories \cite{critique}.
For both the MF-ADH and \expmf theories, these convexity violations occur for additive models ($\delm = -\delp$), roughly, only when $\delm + \rhostar \gtrsim 1$.
No convexity violations are found for $\lambda \lesssim 2$ at any physical density (recalling the packing limit): this confirms the view that the present theories are of greatest validity for modest asymmetries.

\newpage
\section{Predictions for Critical Parameters}
The discussions presented above of the various proposed modifications of the asymmetric DH theory indicate that simpler versions should provide a reasonable basis for further exploration, at least in the case of small size asymmetries.
Accordingly, we present here, first, by way of calibration, the predictions of
(i) the original, 1923 DH theory embodied in (\ref{fbar1923});
(ii) the MF-ADH theory which uses the simple, mean-field border-zone factors (\ref{mfzonefac}); and
(iii) the EXP$^<$MF$^>$ADH theory which retains the mean-field treatment for the super-border zones but recognizes, via (\ref{expzonefac}), the decrease in sub-border zone charge resulting from the Boltzmann-factor enhanced like-charge repulsions.

The predictions for $\tstar_c(\lambda)$ and $\rhostar_c(\lambda)$ for a 1:1 electrolyte are embodied in Fig.\ 4.
The asymmetry variable
$\omega(\lambda) = (1-\lambda)^2 / (1+\lambda^2)$
is convenient \cite{Romero-2000} since it respects the symmetry $\lambda \leftrightarrow 1/\lambda$.
Note that $\omega(2) = 0.20$, $\omega(3) = 0.40$, and $\omega(4.79) = 0.60$, while the point charge limit ($\lambda = \infty$) gives $\omega = 1$.

In Fig.\ 4 no hard-core terms have been included in the free energy \cite{Fisher-Levin}.
Thus for the RPM ($\lambda=1$, $\omega=0$) one has 
${\tstar_c}^{\mathrm{DH}} = \frac{1}{16} = 0.0625$ \cite{fisher,Fisher-Levin}, which may be compared with simulation estimates yielding $\tstar_c \simeq 0.049$ (see \cite{fisher}).
The DH prediction for the critical density is very low, namely, 
${\rhostar_c}^{\mathrm{DH}} = 1/64\pi \simeq 0.00497$ \cite{Fisher-Levin};
however, this increases to around $\rhostar_c \simeq 0.03$ when Bjerrum ion pairing is introduced \cite{fisher,Fisher-Levin}.
Inclusion of hard-core effects, say via a Carnahan-Starling form, reduces all these parameters by a few percent \cite{Fisher-Levin} --- and the same is expected to happen for the ADH-based theories.

We note immediately from Fig.\ 4 that the original DH theory (i) predicts that both $\tstar_c(\lambda)$ and $\rhostar_c(\lambda)$ \emph{rise} rapidly with $\lambda$.
These are precisely the trends found by the MSA (using the energy route) \cite{Gonzalez-1999} and by the MSA with Bjerrum-Ebeling-Grigo pairing \cite{Raineri-2000}.
The modified Poisson-Boltzmann approximations of \cite{sabir} likewise predict that $\rhostar_c$ increases.
(See also Fig.\ 3 in \cite{Romero-2000}.)
In these approximations, however, the initial $\lambda = 1$ values are well known to be significantly higher [ $\tstar_c(1) \simeq 0.08$, $\rhostar_c(1) \simeq 0.015 \mbox{ to } 0.03$ ];
nevertheless, the proportionate rate of increases are roughly comparable.  

By contrast, both of the modified ADH theories predict a strong \emph{decrease} in $\tstar_c$ and $\rhostar_c$ when $\lambda$ increases from unity:
see plots (ii) and (iii).
Furthermore, these decreases are in accord with the simulations \cite{Romero-2000,Yan-2000} (which, however, start from $\tstar_c(1) \simeq 0.049$ and $\rhostar_c(1) \simeq 0.07$) and the relative rates of fall are quite comparable.
For the MF-ADH theory (ii) the critical temperature decreases monotonically and we find $\tstar_c(\lambda = \infty) \simeq 0.049$ in the point charge limit ($\asubpp \app 0$),
while $\rho_c$ exhibits a shallow minimum at $\lambda \simeq 4$ and then increases to $\rhostar_c(\lambda = \infty) \simeq 0.006$.
By comparison, extrapolation of the simulations (beyond $\lambda \simeq 6$) suggests, roughly, $\tstar_c(\infty) \simeq 0.022$ and $\rhostar_c(\infty) \simeq 0.015$ \cite{Romero-2000}.

In the case of the \expmf theory (iii), however, while $\tstar_c(\lambda)$ falls monotonically to about 0.053 when $\lambda \app \infty$, the critical density undergoes a shallow minimum around $\lambda \simeq 1.8$ and then rises.
%, eventually reaching $\rhostar_c(\infty) \simeq XXX$.  
Insofar as the simulations seem trustworthy, and exhibit plots which curve \emph{downwards} (i.e., are concave) vs. $\omega(\lambda)$, it is surprising that the use of the EXP$^<$ choice for $\eless$ leads to apparently inferior predictions.
Indeed, on \emph{a priori} theoretical grounds, the latter would seem superior to the MF assignment $\eless = 1$.
We emphasize again, therefore, that various steps in our analysis appear most soundly based when $\lambda$ is not too large.

For completeness, we report that the MF-ADH and \expmf theories predict that the critical ratio $Z_c = p_c/\rho_c k_B T$ falls monotonically with $\lambda$ from 0.09036 at $\lambda = 1$ \cite{Fisher-Levin}, while the original DH theory (Sec.\ II.C) predicts a fall of just a few percent followed, for $\lambda \gtrsim 2$, by a monotonic rise.
Inclusion of hard-core terms in the free energy increases these values (by about 7 \%) but otherwise the behavior remains similar.

\newpage
\section{Summary and Conclusions}
We have extended \dhtext (DH) theory to asymmetric two-component hard-sphere electrolytes (i.e., ``primitive models'') and computed the predicted critical temperatures and densities.
We also have derived, in Appendix A, a lower bound on the internal energy that extends Onsager's bound \cite{onsag} and depends only on the product of the valences $|\zp \zm|$ and the unlike ``collision diameter'' $\asubpm$.
In order to extend the original DH theory \cite{debye} (based on the Poisson-Boltzmann equation) to the case of size-asymmetric ions with, say, $\asubmm > \asubpp$, we have identified ``border zones'' around ions of both species, which prove to be of essential importance.
These zones are \emph{charge-unbalanced} even at infinite temperature because the larger (negative) ions are geometrically excluded while the smaller (positive) ions may always enter: see Fig.\ 1.

DH extensions which describe the border zones in a physical way (Sec.\ III) prove successful in matching trends --- as determined by two independent simulation studies \cite{Romero-2000,Yan-2000} --- in the critical temperature and density with increasing size asymmetry (see Fig.\ 4 and Sec.\ IV).
This contrasts favorably with other theories, including several based on the mean spherical approximation \cite{sabir,Gonzalez-1999,Raineri-2000}, which predict trends opposite to those revealed unequivocally by the simulations.

The existence of the zones complicates the theory in an essential way;
however, the usual DH approach can be extended straightforwardly and yields explicit approximations for the internal energy (and, thence, results for other thermodynamic properties).
This asymmetric DH (or ADH) theory reproduces the limiting laws and provides the exact high-temperature second-virial coefficients, $B_{2,1}$ and $B_{2,2}$ [see (\ref{fbarexact})-(\ref{b22})], down to the point-ion limit.

However, in contrast to the standard DH theory for the symmetric restricted primitive model (RPM) with $\lambda = \asubmm/\asubpp = 1$, the straightforward ADH theory violates the (extended) lower bound on the internal energy in the coexistence region when $\lambda > 5$. 
Even more seriously, for moderate asymmetries and moderate temperatures, the mean charge in a ``sub-border'' zone (that surrounds a $+$ ion) is predicted to \emph{change sign} and, at low $T$, to diverge;
but, by construction of the model, such behavior is physically impossible!
This pathology is readily traced to use of the standard DH linearization of the Boltzmann factor within the border zones: see Sec.\ II.B.
Modifications of the ADH theory are thus essential for applications at low $T$.
%The resulting two-term expression (\ref{rhoqmid}) --- one of which is linear in $\beta$ --- changes to the wrong sign at low $T$ ($\beta \app \infty$).
%By contrast, in the ``exterior'' zone where ions of both species are present (and the only region outside the central ion for the symmetric case), the linearization causes no harm: 
%the zeroth order terms in $\beta$ cancel one another due to charge neutrality, and the remaining linear terms sum to form the always-positive Debye factor of (\ref{rhoqfar}), which ensures the exterior zone charge will exhibit the appropriate sign for all $T$ --- i.e., opposite to that of the central charge.
%In this light, DH theory for the size-symmetric RPM, which yields sensible low-$T$ results \cite{Fisher-Levin,critique}, owes at least some of its success to cancellations arising from the symmetry.
%This pathology of the ADH theory demands modifications of the usual DH linearization within the border zones.

As shown in Sec.\ III, one may restore physically sensible behavior while retaining the exact high-$T$ behavior by introducing ``border zone factors'', $\eless(T)$ and $\egrtr(T)$, originally proposed in \cite{dmz-thesis}.
(This also ensures that the energy bounds are no longer violated in the critical region and beyond.)
The simplest such modification amounts to a mean-field approach in which the Poisson-Boltzmann factors in the border zones (only) are merely replaced by their $T=\infty$ limits, namely, unity.
For a sub-border zone (around a smaller $+$ ion) a theoretically preferable approach, dubbed EXP$^<$, replaces the self-consistent Poisson-Boltzmann factor by a mean bare-interaction Boltzmann factor [or, better, linearizes about such a mean value: see (\ref{zonefacexplin})].
A corresponding exponential treatment of the super-border zone (around a negative ion) proves, however, more problematical owing to physically crucial charge-saturation effects that are hard to elucidate in a precise way.
One may expect that both the mean-field and EXP$^<$ modifications of linear ADH theory are most reliable at small degrees of asymmetry.

To explore the consequences of the simplest modified mean-field and EXP$^<$ ADH theories (just sketched) we have computed the predicted critical parameters as a function of the size-asymmetry $\lambda$ for 1:1 electrolytes [with additive interactions, $\asubpm = \boxhalf (\asubpp + \asubmm)$]: see Fig.\ 4.
As $\lambda$ increases from unity, the predicted $\tstar_c(\lambda)$ and $\rhostar_c(\lambda)$ \emph{fall} systematically within both of these modified ADH theories.
[See (\ref{rhostar}) and (\ref{tstargen}) for definitions of the reduced units.]
These decreases accord well with recent simulations \cite{Romero-2000,Yan-2000}.
On the other hand, an original proposal by Debye and H\"{u}ckel in 1923, that completely ignores the border zones (see Sec.\ III.C), predicts diametrically \emph{opposite} trends.
Furthermore, current, more sophisticated theories \cite{Gonzalez-1999,Raineri-2000} make similar predictions of increasing $\tstar_c$ and $\rhostar_c$ (in addition to yielding excessively large values of $\tstar_c$ for the RPM \cite{fisher,Fisher-Levin}). 

We conclude that DH-based theories seem to extract the basic physics in a quantitatively more reliable way \cite{fisher,Fisher-Levin}, even for size-asymmetric systems, than do potentially better, but physically less transparent approaches like the MSA.
It is still necessary, however \cite{critique}, to incorporate Bjerrum ion-pairing and dipole-ion solvation \cite{Fisher-Levin} into the modified ADH theories expounded here.
(This will also increase the predicted critical densities to better match the anticipated values.)
It is not obvious that the correct trends with asymmetry (accepting the validity of the simulations \cite{Romero-2000,Yan-2000}) will survive these extensions:
it seems likely, nonetheless, that the proper dependence on asymmetry will be preserved (as suggested, e.g., by comparing the results of \cite{Gonzalez-1999} and \cite{Raineri-2000}).

From a broader perspective, it remains frustrating that more powerful and definitive theoretical techniques have not yet been devised to aid in our understanding of such a fundamental and significant model for condensed matter.

\begin{acknowledgments}
We appreciate the interest, encouragement, and comments of Professors Benjamin P. Vollmayr-Lee, Elliot Lieb, and A. Z. Panagiotopoulos.
We also are grateful to Dr.\ S.\ Banerjee for pointing out an oversight in the appendix.
Professor Katharina Vollmayr-Lee generously translated sections of \cite{debye} related to the present work, and Youngchan Kim kindly commented on a draft typescript.
The bulk of the research reported here was supported by the National Science Foundation (under Grant Nos.\ CHE 96-14495 and CHE 99-81772).
D.M.Z. is grateful for current support from Professor Thomas B.\ Woolf and the National Institutes of Health.
S.B. gratefully acknowledges the current support of Professor Terry Gaasterland.
\end{acknowledgments}

\pagebreak
\appendix
\section{Energy Bounds for the Two-Species Primitive Model}
\label{sec:e2pmbounds}
We here generalize the Onsager bound for the RPM \cite{onsag} to the general two-component hard-sphere model defined in Sec.\ II and also develop an (apparently new) lower bound which improves on the simple Onsager result.

The quantity to be bounded is the {\em interaction} energy of the ions, $\biguint$.
Defining the interionic distances, $r_{ij}$, and interaction potentials, 
\begin{equation}
\label{coulombpotential}
\potij = q_i q_j/D r_{ij},
\end{equation}
where $q_i$ is the charge on the $i$th ion,
 one has
\begin{equation}
\label{biguconfig}
\biguint = {\sum_{i<j}} \potij \; .
%\biguint = {\sum_{i<j}}^{\dagger} \potij, 
\end{equation}
%where the dagger $(\dagger)$ indicates that the interionic distances are constrained by the hard-core conditions embodied in the diameters $\asubpp$, $\asubpm$, and $\asubmm$;
By adopting the $1/r_{ij}$ form in (\ref{coulombpotential}), we implicitly assume that the charge densities between any two ions are spherically symmetric and never overlap. 
To avoid irrelevant singularities, we further assume that the charge density is everywhere finite, i.e., has been suitable ``smeared'' or distributed.
%Note that the interionic distances are constrained by the hard-core conditions embodied in the diameters $\asubpp$, $\asubpm$, and $\asubmm$;
%we shall assume all of these diameters are non-zero.
%It should be recalled that the $1/r_{ij}$ form of the interaction energy terms is reproduced by assuming the non-overlapping condition and that the charge {\em within} any given ion is distributed in a spherically symmetric --- but otherwise arbitrary --- way.
%To avoid irrelevant singularities, we shall assume that the charge is ``smeared'' in some way and not concentrated at a point.
%Note that for an ``attractive'' non-additive SAPM (recall Fig.\ \ref{fig:e2pmattract}) where, say, $\asubpp > \asubpm$ and $\asubmm > \asubmm$, the charge would have to be distributed over a spherical region of diameter $\asubpm$ (less than the like diameters) in order for the $1/r$ potential to hold at all permissible values of $r_{+-}$.

Following Onsager \cite{onsag}, consider the {\em total} electrostatic energy, $\bigutot$, which, besides the interaction energy, also includes the self energies, $\uself_i$, i.e., the energies required to assemble the individual ions from an infinitely dispersed charge state of zero energy.
As is well known (see, e.g., \cite{jackson}), the total energy may be expressed in terms of the electric field, $\mathbf{E}$, as a positive definite quantity, namely,
\begin{equation}
\label{utotesquared}
\bigutot = \sum_{i<j} \potij + \sum \uself_i = \frac{1}{8 \pi} \int \, \dtrstd |{\mathbf{E}}|^2 > 0.
\end{equation}
%Along with (\ref{coulombpotential}), the second equality may be taken as defining the sum of the self energies;
%however, the individual terms, $\uself_i$, are well known \cite{jackson} to correspond to the energies of assembly.
Note that the smeared charge distributions assumed guarantee the finiteness of $\bigutot$.

\subsection{Generalization of the Onsager Bound}
The Onsager bound for the RPM and its direct generalization follow easily from (\ref{utotesquared}).
Rearrangement leads to
\begin{equation}
%\frac{1}{N} \frac{1}{N} 
\sum_{i<j} \potij \geq -\sum_i \uself_i.
\end{equation}
This inequality holds for {\em arbitrary} charge distributions obeying the restrictions stated above;
different distributions, of course, lead to different values of $\uself_i$.
To obtain the strongest bound, we must {\em minimize} the magnitudes of the self energies.
This is accomplished by dispersing the ionic charge as much as possible, namely, by placing it in a thin shell on the surface of the largest permissible sphere (of diameter $\amax$).
Considering a vanishingly thin shell, the minimal self energy is thus
\begin{equation}
\label{uselfoptimal}
\min{\{\uself_i\}} = q_i^2/D \amax_i.
\end{equation}
For the symmetric RPM, the equality of charges, $q_+ = -q_- \equiv q$, and of sizes, $a_i^{\mathrm{max}} \equiv a_{+-} \equiv a$, immediately leads to the Onsager bound $q^2/Da$.
In the case of non-additivity, however, one must avoid overlapping charge distributions from distinct ions, so that
\begin{equation}
\label{onsagerdiameter}
\amax_i = \min{\{\asubpm,a_{ii}\}} \; .
\end{equation}

We can now formulate the explicit Onsager bound.
First, defining $z = |\zp/\zm|$, note that the total numbers of particles of each species are
\begin{equation}
\label{nplus}
N_+ = N/(1+z)
\mbox{ \hspace{0.5cm} and \hspace{0.5cm} }
N_- = zN/(1+z) 
\;.
\end{equation}
Recalling the definition of the reduced energy (\ref{ue2pm}) and using the constraint of overall charge neutrality, one finds the bound, namely,
\begin{equation}
\label{onsagere2pm}
u(\rho,T) \geq u_{\mathrm{Ons}} 
  = - \frac{z(a/\amax_{+}) + (a/\amax_{-})}{1+z} \;.
%
%  = - \lpar \frac{z(a/\amax_{+})}{1+z} 
%	+ \frac{(a/\amax_{-})}{1+z} \rpar.
%
%  = - \lpar \frac{|\zp/\zm|(a/\amax_{+})}{1+|\zp/\zm|} 
%	+ \frac{|\zm/\zp|(a/\amax_{-})}{1+|\zm/\zp|} \rpar.
%
%  = - \frac{1}{2} \lpar \frac{|\zp/\zm|}{\amax_{+}/a} 
%	+ \frac{|\zm/\zp|}{\amax_{-}/a} \rpar.
\end{equation}
Note that for the size-symmetric case, where $\amax_{+}=\amax_{-}=a$, the charge asymmetry does not affect the bound in these reduced units, so that 
$u(\rho,T) \geq u_{\mathrm{Ons}} = - 1$.
If we separate the three basic size asymmetry cases,
% and employ the ratio $z = |\zp/\zm|$, 
the result (\ref{onsagere2pm}) translates into
\begin{equation}
\begin{array}{rll}
\label{onsagcases}
u_{\mathrm{Ons}}^{\irm} & 
  = - \{ z / [(1 + \delta_+)(1+z)] 
	+ 1 / [(1 + \delta_-)(1+z)] \} \; ,  
%
%  = - \frac{1}{2} \lbrak |\zp/\zm| / (1 + \delta_+) 
%	+ |\zm/\zp| / (1 + \delta_-) \rbrak \; ,  
  %= - \frac{1}{2} \lpar \frac{|\zp/\zm|}{\asubpp/a} 
	%+ \frac{|\zm/\zp|}{\asubmm/a} \rpar 
        \hspace{2em}
& \mbox{for } \asubpp , \; \asubmm < a \\
u_{\mathrm{Ons}}^{\iirm} & 
  = - \{ z / [(1 + \delta_+)(1+z)] 
	+ 1/(1+z) \} \; , 
%
%  = - \frac{1}{2} \lbrak |\zp/\zm| / (1 + \delta_+) 
%	+ |\zm/\zp| \rbrak \; , 
  %= - \frac{1}{2} \lpar |\zp/\zm| 
	%+ \frac{|\zm/\zp|}{\asubmm/a} \rpar 
& \mbox{for } \asubpp < a < \asubmm \\
u_{\mathrm{Ons}}^{\iiirm} 
& = - 1 \; , 
%& = - \frac{1}{2} \lpar |\zp/\zm| + |\zm/\zp| \rpar \; , 
& \mbox{for } a < \asubpp, \; \asubmm \; .
\end{array}
\end{equation}

Note that the strongest bound is 
%$-\frac{1}{2}(|\zp/\zm| + |\zm/\zp|)$, 
-1,
which obtains for Case III, when the like-diameters both exceed $a$,
%this translates into a bound of $-1$ for a 1:1 electrolyte, 
which matches the RPM result.
On the other hand, $u_{\mathrm{Ons}}$ is weaker in Cases I or II,
% i.e., if either (or both) of the like-diameters are {\em smaller} than $\asubpm \equiv a$.
%Indeed, in these cases, 
since the bound can then diverge to $-\infty$ when $\delta_\sigma \app -1$ (or, equivalently, as $a_{\sigma\sigma} \app 0$). 
Physically, shrinking the like-diameters (but keeping $\asubpm$ positive) should not decrease the energy:
thus a stronger bound is desirable.

\subsection{Improved Bound}
To do better we compare the two-species primitive model with another model whose energy can be bounded by an Onsager construction, specifically, an interpenetrating, two-species \emph{shell} model, consisting of a charge-neutral mixture of uniformly surface-charged spheres of total charges $q_i = \zp \elcharge$ or $\zm \elcharge$ and equal diameters $a$, but with {\em no} hard-core constraint.
The shell diameter $a$ will be identified with $a_{+-}$ for the primitive models.
The interaction potential is described further below.

Since the configuration space of the primitive model --- in terms of ion-center locations --- is a subset of the space of the shell model, any ground state configuration of the primitive model of energy, say $\biguint_0$, is present in the shell model with energy, say $\biguint_1$, which cannot be lower than the shell model ground state, say $\biguint_{0,\mathrm{S}}$.
Thus, if we can establish
\begin{equation}
\label{isitmore}
\biguint_0 \geq \biguint_1
\end{equation}
and also show that $\biguint_{0,\mathrm{S}}$ is bounded below, we will obtain a lower bound on $\biguint$, as desired.

We may construct the interaction potential of the interpenetrating shell model, say $\potijtwid$ (which does not behave as $1/r_{ij}$ at all separations), as the difference between the total energy of a two-particle system and the sum of the two self energies.
Thus if $\rhohat_k(\rbf;\rbf_k)$ represents the charge distribution of a shell ion $k$ centered at $\rbf_k$, and
\begin{equation}
\label{chargeadd}
\rhohat({\mathbf{r}}) \equiv \rhohat_i({\mathbf{r}};\rbf_i) + \rhohat_j({\mathbf{r}}; \rbf_j)
\end{equation}
is the total charge density for two ions, the interaction potential follows from
\begin{equation}
\label{phirism}
\label{phisrismesqared}
\potijtwid + \frac{q_i^2}{Da} + \frac{q_j^2}{Da} 
  = \frac{1}{8\pi} \int \dtrstd |{\mathbf{E}}|^2 
	= \boxhalf \int \dtrstd \rhohat({\mathbf{r}}) \phi({\mathbf{r}}) \; ,
\end{equation}
where ${\mathbf{E}}(\rbf; \rbf_i, \rbf_j)$ and $\phi(\rbf; \rbf_i, \rbf_j)$ are the total field and electrostatic potential.
If we now define
$\phi_k({\mathbf{r}}; \rbf_k)$ to be the electrostatic potential resulting from an isolated shell ion $k$ at $\rbf_k$, we have
\begin{equation}
\label{uselfcorrect}
\boxhalf \int \dtrstd \rhohat_k({\mathbf{r}}; \rbf_k) \phi_k({\mathbf{r}}; \rbf_k) 
  = \frac{q_k^2}{Da} \; .
\end{equation}
Then, using the linearity of the charge density and potential, the relation (\ref{phirism}) may be simplified to yield
\begin{equation}
\label{phirismint}
\pottwid_{ij} \equiv \pottwid(r_{ij}) = \boxhalf \int \dtrstd 
	\lbrak \rhohat_i({\mathbf{r}}; {\mathbf{r}}_i) \phi_j({\mathbf{r}};\rbf_j) 
		+ \rhohat_j({\mathbf{r}}; {\mathbf{r}}_j) \phi_i({\mathbf{r}};\rbf_i)
			 \rbrak .
\end{equation}

To compare the energy of an arbitrary primitive-ion configuration with that of the corresponding shell-ion configuration in order to establish (\ref{isitmore}), we observe first that because of the pairwise additivity of the interaction energy (\ref{biguconfig}), one need analyze only two shell ions of the same charge which \emph{overlap}.
To see this, note that all $(+,-)$ ion pairs in a primitive model are separated by distance not less than $\asubpm = a$, which is the same diameter as the shell ions.
Thus, oppositely charged shell ions that correspond to a primitive ion configuration never overlap.
The only differences arise when, in the primitive ion system, one or both of the like diameters, $a_{\sigma \sigma}$, is smaller than $a$.
This will allow overlapping shells in the corresponding shell configurations.
If $\asubpp$ and $\asubmm$ exceed $a$, the energy of corresponding primitive and shell configurations will always be identical.

Consider then, for concreteness, two positive overlapping shells,
separated by a distance $r_{++}$, with $\asubpp < r_{++} < a$;
see Fig.\ 5.
We want to show that the interaction energy of this pair, $\pottwid(r_{++})$, is less than that for the the corresponding non-overlapping primitive ions which is simply $q_+^2/Dr_{++}$.

By symmetry the two terms in (\ref{phirismint}) are now identical.
Thus, consider the charge distribution of the right-hand shell in Fig.\ 5 in the potential of that on the left;
the right-hand distribution divides naturally into the two parts shown in the figure: a part exterior to the left-hand shell (bold) and an interior part (dashed).
If $\rbf_l$ denotes the position of the left-hand ion, the resulting potential at an exterior points, $\rbf^>$, is simply $q_+/|\rbf_l - \rbf^>|$ (because these points ``see'' a spherically symmetric left-hand charge distribution).
Conversely, interior points, such as $\rbf^<$, experience only the {\em constant} electrostatic potential $q_+/(a/2)$. 
This is clearly {\em less than} the potential they would experience were all the left-hand charge distributed on the smaller primitive ion sphere with the same center, $\rbf_l$.
Consequently, overlapping like-charged shell ions have a {\em smaller} interaction potential than the corresponding primitive ions with the same centers.
This establishes (\ref{isitmore}).

To obtain a lower bound for the shell model itself, recall (\ref{utotesquared}) and (\ref{chargeadd}) and bound the total energy of any shell configuration as
\begin{equation}
\label{utottilesquared}
\bigutot_{\mathrm{S}} = \frac{1}{8\pi} \int \dtrstd |{\mathbf{E}}|^2 = \boxhalf \int \dtrstd \rhohat({\mathbf{r}}) \phi({\mathbf{r}}) > 0 \; ,
\end{equation}
where the total shell charge density, $\rhohat(\rbf)$ can be expressed in the form (\ref{chargeadd}) but with a sum extending over all the shell ions.
The total electrostatic potential, $\phi(\rbf)$, can be decomposed similarly, yielding
\begin{equation}
\label{genladd}
\boxhalf \!\! \int \dtrstd \rhohat({\mathbf{r}}) \phi({\mathbf{r}})
= \boxhalf \!\! \int \dtrstd 
	\lbrak \rhohat_1({\mathbf{r}}) +
	%\rhohat_2({\mathbf{r}}) + 
        \cdots +
	\rhohat_N({\mathbf{r}})\rbrak
	\lbrak \phi_1({\mathbf{r}}) +
	 %\phi_2({\mathbf{r}}) + 
         \cdots +
	 \phi_N({\mathbf{r}}) \rbrak.
\end{equation}
Finally, by combining the previously defined shell self energies, (\ref{uselfcorrect}), and interaction potentials in (\ref{phirismint}), the inequality (\ref{utottilesquared}) may be re-arranged to give
\begin{equation}
\label{boundrism}
\frac{ \biguint_{\mathrm{S}} }{N} = \frac{1}{N}\sum_{i<j} \potijtwid 
  \geq \frac{ \biguint_{0,\mathrm{S}} }{N}
  \geq -\frac{1}{N}\lpar \frac{N_+ q_+^2}{Da} + \frac{N_-q_-^2}{Da} \rpar
  = -\frac{|q_+ q_-|}{Da} \;.
%  \geq -\boxhalf \lpar \frac{q_+^2}{Da} + \frac{q_-^2}{Da} \rpar.
\end{equation}

The bound on the primitive model is now completed by combining this with (\ref{isitmore}).
In terms of the reduced energy per particle, (\ref{ue2pm}), the result may be written
\begin{equation}
\label{boundatlast}
u(\rho,T) \geq -1 \; .
%u(\rho,T) \geq -\boxhalf \lpar |\zp/\zm| + |\zm/\zp| \rpar \; .
\end{equation}
Note that the positive definite collision diameter, $a_{+-} \equiv a$, and the valences $z_\sigma$ do not appear explicitly here since they enter into the definition (\ref{ue2pm}).
Since $\asubpp$ and $\asubmm$ are also absent, the bound remains valid for point ions ($a_{\sigma\sigma} \app 0$).

\newpage
\begin{center} \Large Figure Captions
\end{center}

\noindent 1. Illustration of a border zone.
The grey sub-border zone may be occupied only by the {\em centers} of $+$ ions since $\asubpp < \asubpm$.

\vspace{0.5cm}

\noindent 2. Effects of size asymmetry on high- and low-temperature energy isochores, according to the linear ADH theory for a 1:1 primitive model.
For each density, the RPM isochores (with $\lambda\equiv\asubmm/\asubpp = 1$) are shown as solid curves.
The associated, successively lower plots correspond to $\lambda = 2$ (dashed), 4 (dot-dashed), and 6 (dotted, for $\rhostar = 0.1$ only). 
The density-dependent, $\tstar \app \infty$ limiting values are {\em exact}.
(The near-agreement between two of the curves at $\tstar = \infty$ is a coincidence.)

\vspace{0.5cm}

\noindent 3. Effects of size asymmetry on the low-temperature energy isochores, according to the simplest ``mean-field'' modification of ADH theory for a 1:1 primitive model at densities $\rhostar = 0.01$, 0.03, and 0.1 and size asymmetries
$\lambda = 1$ (the RPM) solid curves,
$\lambda = 2$, dashed, and
$\lambda = 4$, dot-dashed curves.
%Compare to the linear theory in Fig.\ \ref{fig:uladh}(b), noting especially the difference in vertical scales.
%In the nonlinear theory, no bound violation occurs and the low-temperature energy values of the asymmetric models \emph{exceed} that of the symmetric RPM.

\vspace{0.5cm}

\noindent 4. Critical temperature and density predictions for a 1:1 electrolyte with additive hard-sphere interactions as a function of the size asymmetry variable
$\omega(\lambda) = (1-\lambda)^2 / (1+\lambda^2)$, 
which increases monotonically with $\lambda = \asubmm / \asubpp$.
The reduced parameters $\rhostar_c$ and $\tstar_c$ are defined via (\ref{rhostar}) and (\ref{tstargen}):
(i) represents the 1923 DH theory, while modifications of asymmetric DH theory are embodied in
(ii) with mean-field factors for both $\eless$ and $\egrtr$, and
(iii) with a Boltzmann-factor (EXP$^<$) border zone factor for $\eless$ and a mean-field factor $\egrtr$.
The circles denote the $\tstar_c$ simulation estimates of Romero-Enrique \emph{et al.}\ \cite{Romero-2000}.

\vspace{0.5cm}

\noindent 5. Two overlapping shell ions, with corresponding hard-core primitive ions (dotted), which are smaller but concentric.
The shell diameter, $a$, is identified with $\asubpm$ of the primitive models.

\newpage

\newpage

\newpage
\begin{figure}
\centerline{ \epsfxsize = 5in
%\epsfclipon
\epsffile{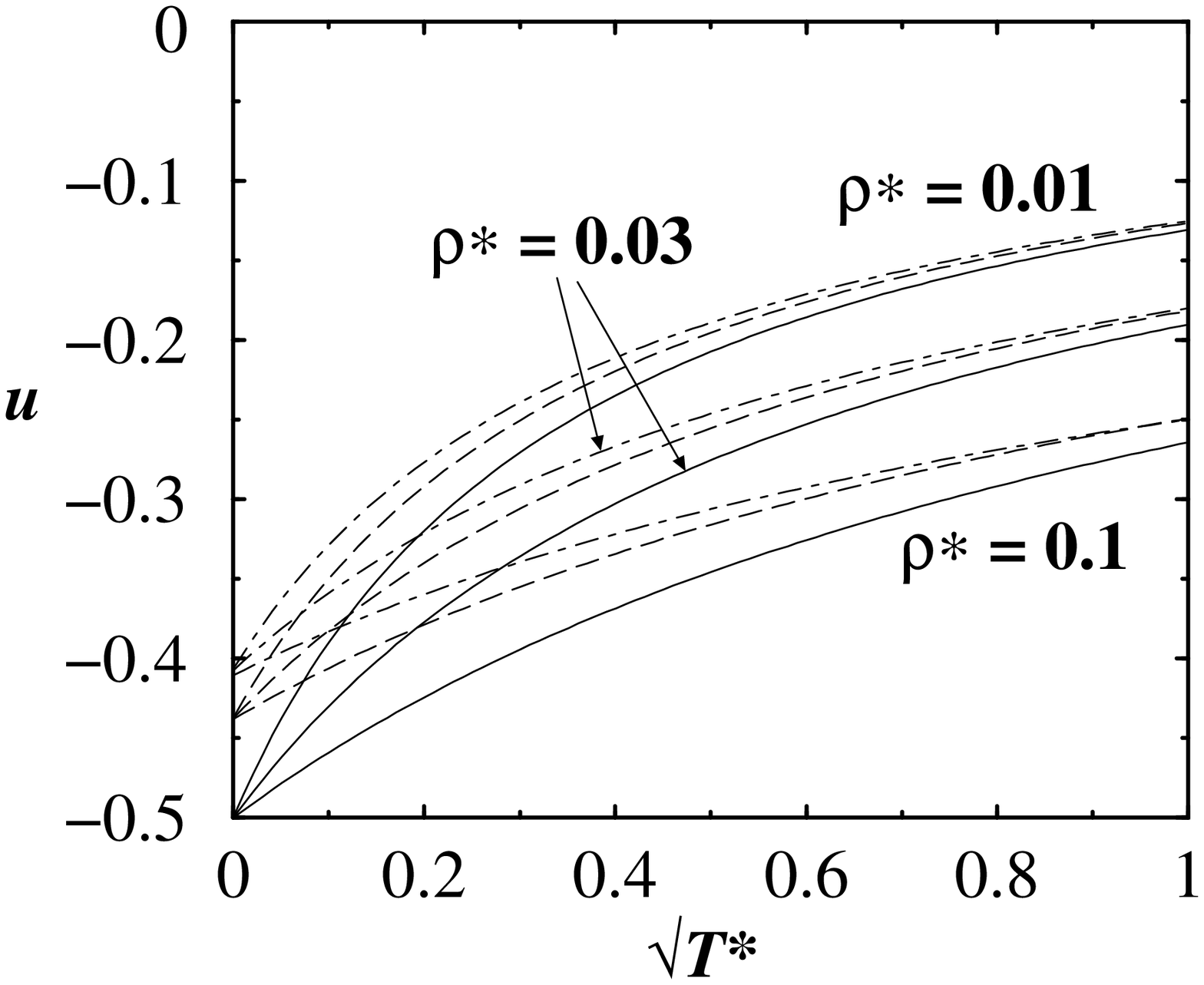}
}
\vspace*{10cm}
\caption
{\label{fig:unladh}
Zuckerman, Fisher, and Bekiranov
}
\end{figure}

\newpage
\begin{figure}
\hspace*{-5.6mm}
\centerline{ \epsfxsize = 4.68in
\epsffile{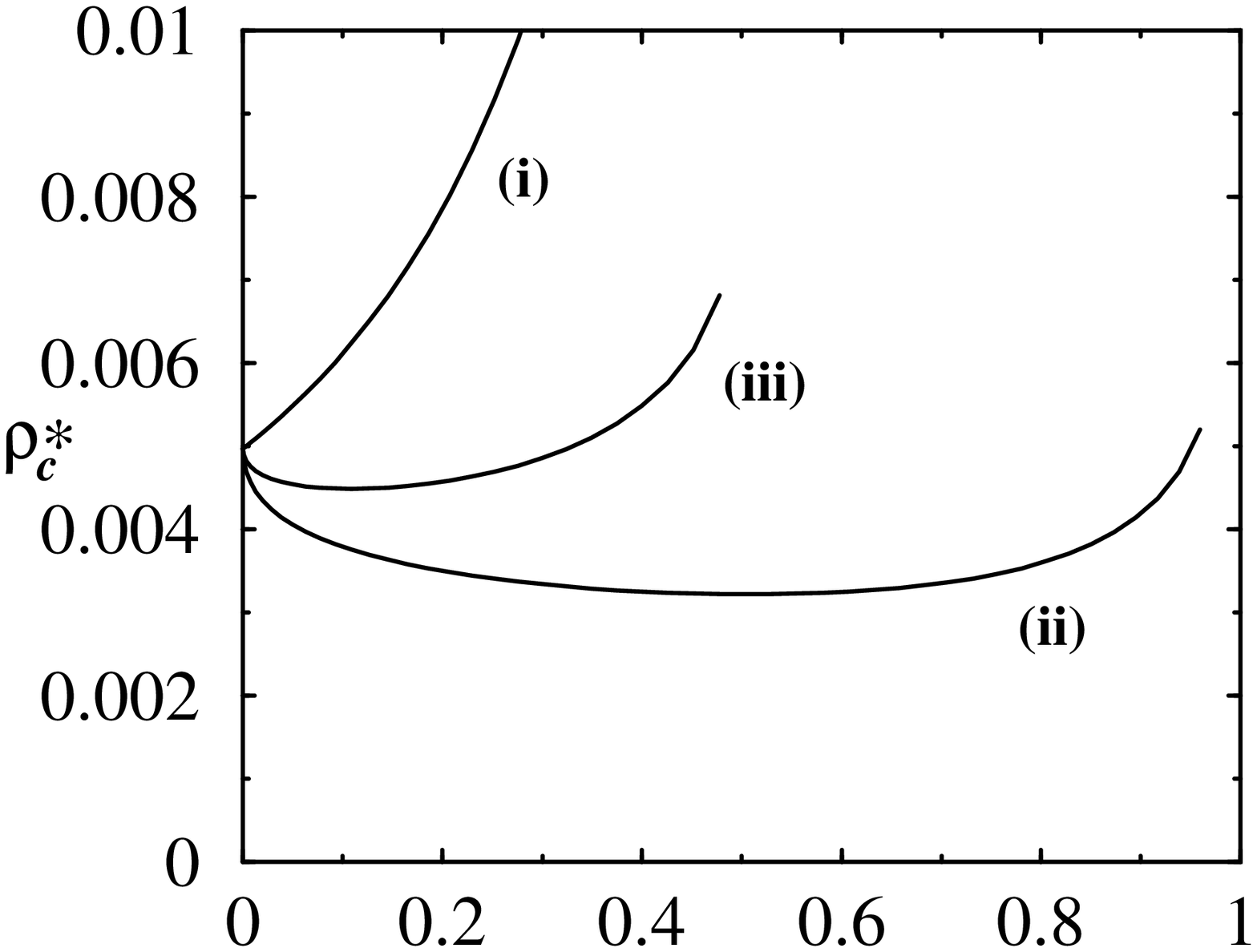}
}
\centerline{ \epsfxsize = 4.8in
\epsffile{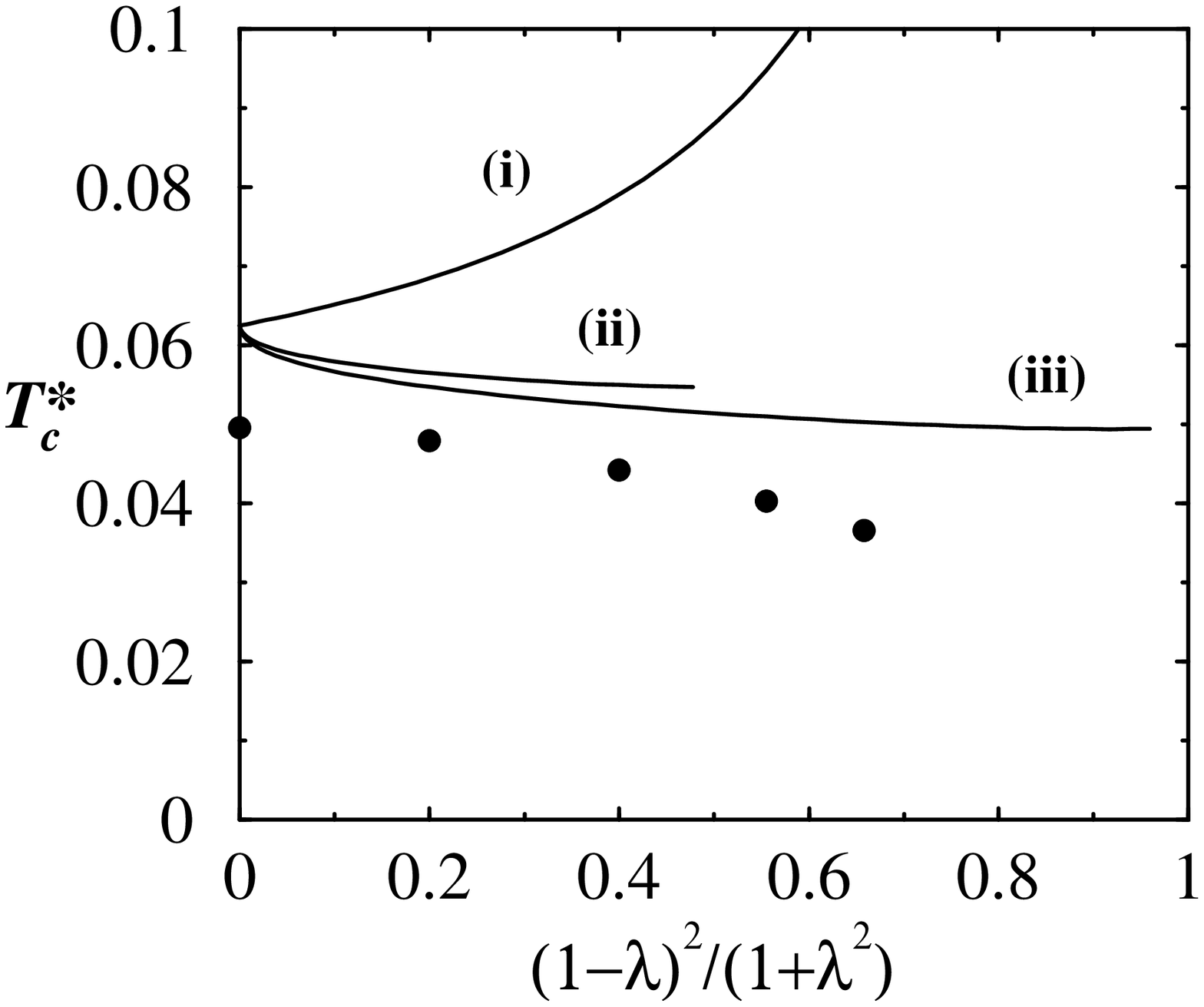}
}
\vspace*{2cm}
\caption
{\label{fig:critnladh}
Zuckerman, Fisher, and Bekiranov
}
\end{figure}

\newpage
\begin{figure}
\centerline{ \epsfxsize = 5.5in
%\epsfclipon
\epsffile{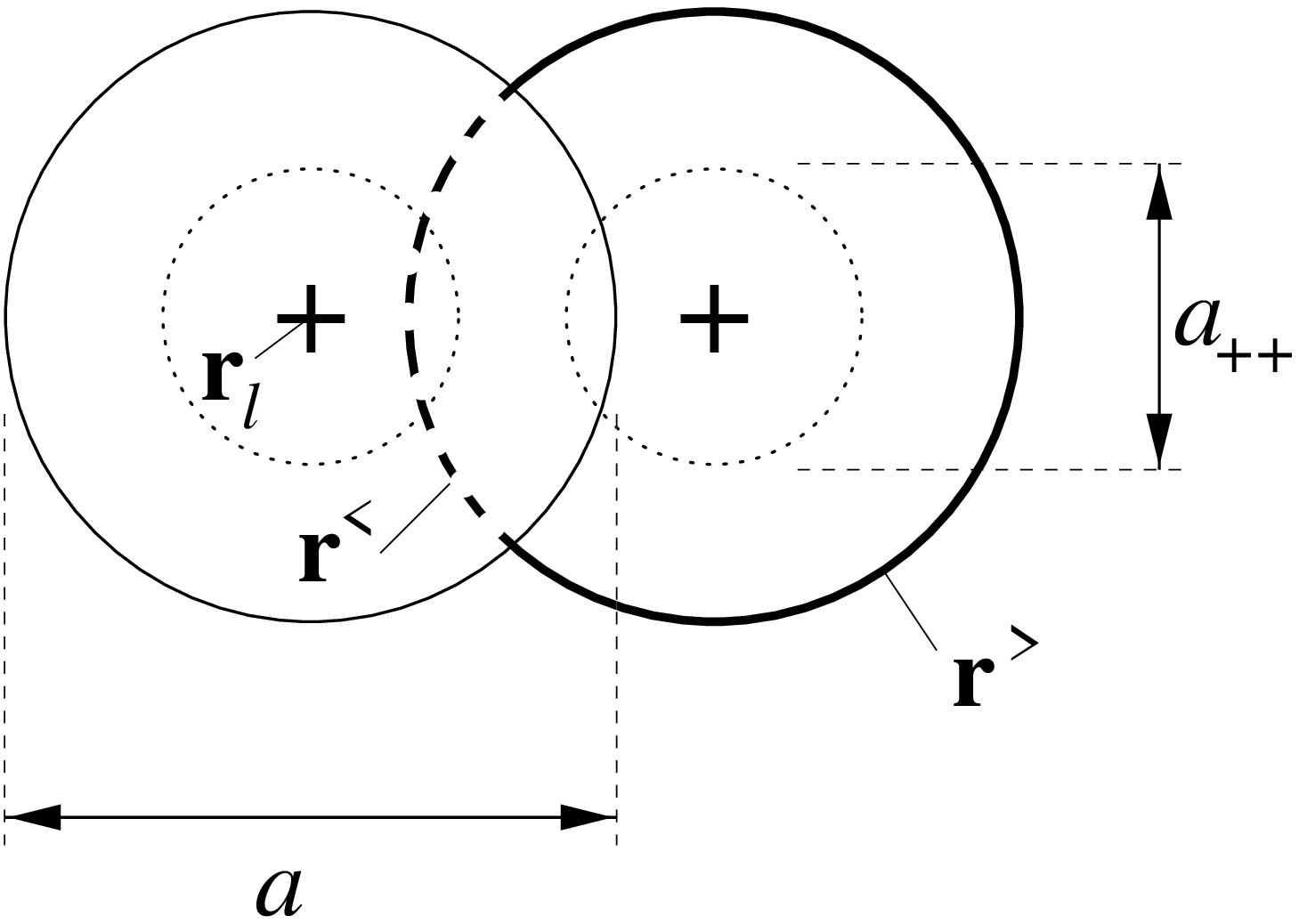}
}
\vspace*{10cm}
\caption
{\label{fig:e2pmoverlap}
Zuckerman, Fisher, and Bekiranov
}
\end{figure}

\end{document}